\documentclass[a4paper,pra,aps,twocolumn,amsfonts,amsmath,amssymb,superscriptaddress,nofootinbib]{revtex4}

\usepackage{graphicx}
\usepackage{dcolumn}
\usepackage{bm}
\usepackage{bbm}
\usepackage{mathtools}
\usepackage{physics}
\usepackage{epsfig}
\usepackage{verbatim}
\usepackage[bottom=0.8in,top=1in,left=0.8in,right=0.8in]{geometry}
\usepackage{units}
\usepackage{esint}
\usepackage[dvipsnames]{xcolor}
\usepackage{hyperref}
\hypersetup{colorlinks=true,
urlcolor={blue},linkcolor={blue},
citecolor={blue}
}
\usepackage{placeins}
\usepackage{microtype}
\usepackage{xspace}

\begin{document}


\def\ii{{\rm i}}  \def\ee{{\rm e}}
\def\rb{{\bf r}}  \def\Rb{{\bf R}}  \def\ub{{\bf u}}  \def\vb{{\bf v}}
\def\xx{\hat{\bf x}}  \def\yy{\hat{\bf y}}  \def\zz{\hat{\bf z}}  \def\nn{\hat{\bf n}}  \def\rr{\hat{\bf r}}
\def\kb{{\bf k}}  \def\kpar{k_\parallel}  \def\kparb{{\bf k}_\parallel} \def\qparb{{\bf q}_\parallel} \def\Gparb{{\bf G}_\parallel}
\def\Qb{{\bf Q}}  \def\Gb{{\bf G}}  \def\gb{{\bf g}} \def\qx{q_x} \def\qy{q_y}
\def\me{m_{\rm e}}  \def\kB{{k_{\rm B}}}
\def\Eb{{\bf E}}  \def\Bb{{\bf B}}  \def\Ab{{\bf A}}
\def\pb{{\bf p}}  \def\jb{{\bf j}}
\def\rp{r_{\rm p}}  \def\rs{r_{\rm s}}
\def\vF{v_{\rm F}}  \def\kF{{k_{\rm F}}}  \def\EF{{E_{\rm F}}}
\def\blue{\color{blue}}  \def\red{\color{red}}  \def\green{\color{green}}
\def\lamp{\lambda_{\rm p}}  \def\Qp{Q_{\rm p}}  \def\Lp{L_{\rm p}}  \def\kp{k_{\rm p}}  \def\wp{\omega_{\rm p}}
\def\Htot{\hat{\mathcal{H}}_{\rm tot}} \def\Hint{\hat{\mathcal{H}}_{\rm int}} 
\def\Heint{\hat{\mathcal{H}}_{\rm int}^{\rm eff}}
\def\hr{\hat{\rho}}
\def\Hzu{\hat{\mathcal{H}}^{\rm univ}_0} \def\hr{\hat{\rho}}
\def\Hze{\hat{\mathcal{H}}^{\rm e}_0} \def\hr{\hat{\rho}}
\def\hjb{\hat  {\bf j}} \def\hAb{ \hat{\bf A}} \def\ha{{\hat a}} \def\hb{{\hat b}} \def\hS{{\hat{\mathcal{S}}}}
\def\qb{{\bf q}} \def\hU{\hat{\mathcal{U}}}
\def\PM{{P\!M}} \def\fb{{\bf f}} \def\zb{{\bf z}} \def\qb{{\bf q}}
\def\mb{{\bf m}} \def\sb{{\bf s}} \def\yb{{\bf y}} \def\hn{{\hat{n}}} \def\hE{{\hat{E}}}

\def\mb{{\bf m}} \def\sb{{\bf s}} \def\yb{{\bf y}} \def\ha{{\hat a}} \def\hb{{\hat b}} \def\hS{{\hat{\mathcal{S}}}}
\def\hc{{\hat c}} \def\NA{{\rm NA}} \def\hn{{\hat{n}}}

\newcommand{\MPINAT}{Department of Ultrafast Dynamics, Max Planck Institute for Multidisciplinary Sciences, D-37077 G\"{o}ttingen, Germany}
\newcommand{\UGOE}{4th Physical Institute - Solids and Nanostructures, Georg-August-Universit\"{a}t G\"{o}ttingen, D-37077 G\"{o}ttingen, Germany}

\title{Tunable quantum light by modulated free electrons}
\author{Valerio Di Giulio}
\email{valerio.digiulio@mpinat.mpg.de}
\affiliation{\MPINAT}
\affiliation{\UGOE}
\author{Rudolf Haindl}
\affiliation{\MPINAT}
\affiliation{\UGOE}
\author{Claus Ropers}
\affiliation{\MPINAT}
\affiliation{\UGOE}
\begin{abstract}
Nonclassical states of light are fundamental in various applications, spanning quantum computation to enhanced sensing. Fast free electrons, which emit light into photonic structures through the mechanism of spontaneous emission, represent a promising platform for generating diverse types of states. Indeed, the intrinsic connection between the input electron wave function and the output light field suggests that electron-shaping schemes, based on light-induced scattering, facilitates their synthesis. In this article, we present a theoretical framework capable of predicting the final optical density matrix of a generic $N$-electron state that can also account for post-sample energy filtering. By using such framework, we study the modulation-dependent fluctuations of the $N$-electron emission and identify regions of Poissonian and super-Poissonian statistics. In the single-electron case, we show how coherent states with nearly $90\%$ purity can be formed by pre-filtering a portion of the spectrum after modulation, and how non-Gaussian states are generated after a precise energy measurement. Furthermore, we present a strategy combining a single-stage electron modulation and post-filtering to harness tailored light states, such as squeezed vacuum, cat, and triangular cat states, with fidelity close to $100\%$.
\end{abstract}
\maketitle

\section{Introduction}
Fast electrons in scanning and transmission electron microscopes (SEM/TEM) offer the capability to measure different material properties with nanometer resolution, thanks to their exceptionally small wavelength. For instance, inelastically scattered electrons carry information about the excitations of a sample, such as phonons \cite{KLD14,HKR19}, plasmonic resonances \cite{proc048,HDK09,DFB12,KGS18}, and geometrically confined dielectric modes \cite{paper361,AFD21}, which can be retrieved by analyzing their final spectrum through electron energy-loss spectroscopy (EELS) \cite{paper149,KDH19}. 

In the past two decades, efforts to improve the spectral resolution, limited in EELS measurements due to the broad-band nature of fast charged particles \cite{paper149}, and to achieve time-resolved imaging, have led to the integration of optical systems into TEM. In such instruments, a laser and an electron pulse interact at the sample, resulting in inelastic electron-light scattering (IELS) \cite{BFZ09,PLZ10}. In the form of photon-induced near-field electron microscopy (PINEM), this combination of techniques has produced remarkable results in studying the femtosecond dynamics of near fields carried by polaritons in nanostructures \cite{HRF21,KDS21,paper419,paper431,MLS24} and optical nonlinearities in dielectric resonators \cite{WHR24}. Beyond imaging, IELS has proven to be an important phenomenon for coherently shaping the longitudinal \cite{FES15} and transverse \cite{paper332,FKN24} full three-dimensional wave function of an electron beam (e-beam). In this context, a general IELS interaction with laser frequency $\omega_{\rm L}$ near a plane positioned at $z$ along the propagation axis, brings an electron traveling with velocity $v$ into the superposition state \[\psi_e(z)=\psi_0(z) \!\! \sum_{\ell=-\infty}^\infty \!\!c_\ell \,\ee^{\ii \ell \omega_{\rm L} z /v}\] composed of momentum coefficients $c_\ell$ and an envelope $\psi_0(z)$. Controlling the amplitude and phase of these coefficients is crucial for attosecond bunching of the electron density \cite{PRY17,KSH18,MB18_2}. Several schemes combining multiple IELS interaction zones have been proposed \cite{YFR21,paper415} to achieve extreme temporal compression, including the replacement of laser illumination with a quantum light source \cite{paper360}.  

Free electrons in SEM/TEM also represent a unique platform for tailoring and probing quantum characteristics of polaritonic modes, either confined, or guided within photonic structures \cite{paper339,KTY23,AF24}. In the case of bosonic statistics, it was shown that there exists a direct relationship connecting the incoming electron coefficients $c_\ell$ and the output mode density matrix $\rho_p$ \cite{paper360,HRN21}, thus rendering a tailored IELS modulation an excellent mean to control $\rho_p$. Under the usual conditions of electron-light coupling linear in the electric field of the mode \cite{K19,paper339}, Poissonian-distributed emission is predicted by single electrons, with a state purity determined by the temporal structure of the electron density \cite{paper360}. Since a possible way of generating quantum light exploits a nonlinear interaction, schemes based on quadratic ponderomotive coupling to generate squeezing \cite{paper402} or incorporating final electron energy filtering (post-filtering) have been proposed \cite{paper180,HRN21} and applied to the generation of few-photon Fock states \cite{FHA22,AHF24}. Furthermore, more complex light states, such as cat and Gottesman-Kitaev-Preskill (GKP) states \cite{GKP01}, were shown to be producible by employing multiple electrons shaped into idealized electron superpositions, characterized by $c_\ell$ with constant phase and amplitude at all orders $\ell$ \cite{DBG23}. 

This article aims at exploring in detail the connection between electron energy modulation and light emission in a single photonic mode with a particular focus on quantum light synthesis. The work is organized as follows. In Sec. \ref{seca}, we develop a general theoretical framework for a linear type -- with an interaction Hamiltonian proportional to the mode electric field -- of electron-light coupling capable of connecting, through an input-output relation, an incoming $N$-electron density matrix with $\rho_p$. In addition, the action of an electron spectrometer is incorporated in the theory to account for the possibility of energy post-filtering. We predict a super-Poissonian light emission for most electron modulations and Poissonian statistics in specific limiting cases when no post-filtering is performed. In Sec. \ref{secb}, we analyze, for single-electron pulses, the coherence conditions and the corresponding modulation requirements to generate high-purity states, both with and without post-filtering. By focusing on the latter case, we propose a simple modulation scheme that combines a strong IELS interaction with an energy filter placed before the sample to significantly enhance electron coherence and state purity. Moreover, for electron ensembles with coherence times longer than the optical cycle of the mode and incorporating post-filtering, we show that pure light states are produced regardless of the form of $c_\ell$. In Sec. \ref{secc}, we leverage the implications of the previous result to explore how a standard IELS modulation can create cat states. Subsequently, in Sec. \ref{secd}, we adopt an approach used for electron-pulse shaping \cite{paper415} combined with an optimization algorithm to provide specific guidelines for designing near-field distributions to be used in an IELS interaction with a laterally patterned nanostructure. We find that squeezed vacuum, cat, and triangular cat states can also be generated with $\sim 100\%$ fidelity under strong coupling conditions and with modulation parameters accessible to state-of-the-art setups. Finally in Sec. \ref{secz}, we discuss the results, their possible extensions, and we provide considerations on the application of the proposed strategies. 

In addition to their theoretical significance, our results represent a fundamental step towards developing practical methods for harnessing nonclassical light from free electrons.

\section{Results and discussion}
\subsection{Output light density matrix after interaction with $N$ electrons \label{seca}}
\begin{figure}
\includegraphics[scale=0.92]{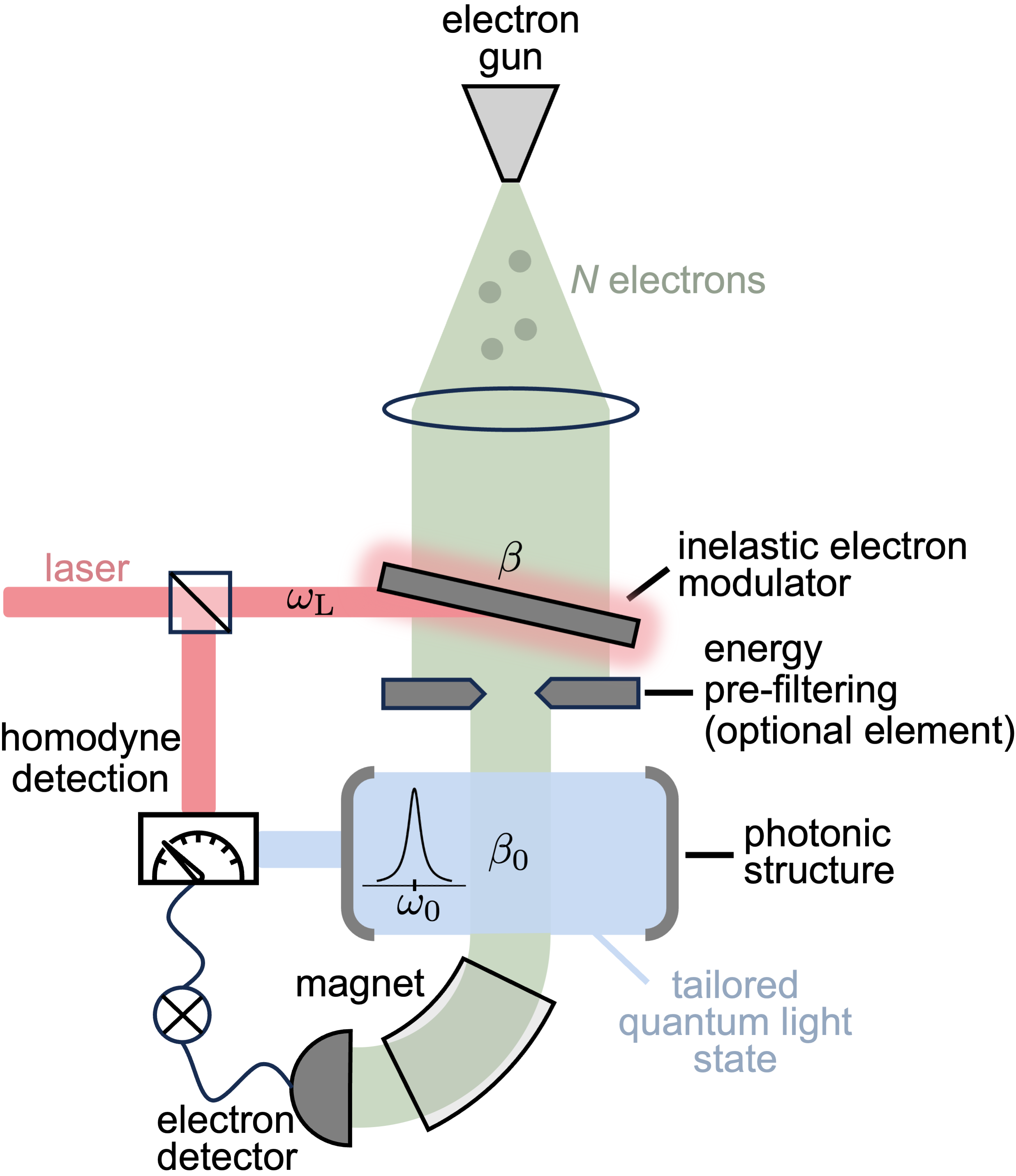}
\caption{{\bf Creation and analysis of quantum light states generated by free electrons.} An e-beam pulse composed by $N$ electrons is  directed into a light-based inelastic modulator that coherently reshapes the electron energy distribution through a single IELS interaction with coupling coefficient $\beta$ and frequency $\omega_{\rm L}$. An optional energy filter placed before the sample may eliminate electrons outside a selected energy range. The beam subsequently passes a nanostructure and emits photons into an optical mode with frequency $\omega_0$ via spontaneous emission of strength $\beta_0$. After this interaction, the generated light is extracted from the structure, and its quantum state is analyzed using a homodyne detection scheme in coincidence with the energies measured by an electron spectrometer composed by a sector magnet and an electron detector.}
\label{Fig1}
\end{figure}

In this work, we study the quantum properties of light emitted by the interaction of an e-beam at kinetic energies in the keV range with a single optical mode of energy $\hbar \omega_0$ and an electric field profile $\vec{\mathcal{E}}_0(\rb)$ in a photonic structure. In particular, we are interested in computing the post-interaction light density matrix $\rho_p$ for electrons having passed through a modulation stage that may comprise an IELS interaction and an energy filter before the sample (pre-filtering). Moreover, we consider the consequences linked to light generation when only a subset of events, determined by a particular choice of the electrons' final energies, is considered (post-filtering) [see Fig. (\ref{Fig1})]. In doing so, we will assume each e-beam pulse to contain $N$ electrons, all with central velocity $\vb=v\zz$, that corresponds to a kinetic energy $E^e_0 \gg \hbar \omega_0 $, and to be well-focused around the transversal coordinate $\Rb$. 

Under these conditions, the quantum evolution of the joint electron-light state can be written by linearizing the electron dispersion, directly leading to the closed form of the scattering operator $\hS=\ee^{\ii \hat{\chi} }\,\hU$ (see Appendix \ref{secA}), with
\begin{align}
\hU=\ee^{\beta_0(\hb \ha^\dagger -\hb^\dagger \ha)}\label{scatt},
\end{align}
written in terms of electron $\hb^\dagger$, $\hb$ and photon $\ha^\dagger,\ha$ creation and annihilation operators. While $\ha,\ha^\dagger$ act on the number of photons, subtracting and adding one particle, respectively, $\hb$ decreases and $\hb^\dagger$ increases the longitudinal  momentum of each electron of $\omega_0/v$. In particular, the former follow boson statistics, whereas, in the considered no-recoil approximation, which is very well justified at high electron energies, the latter commute $[\hb,\hb^\dagger]=0$ \cite{FES15,K19,paper373}. The coupling coefficient 
$\beta_0=(e/\hbar \omega_0)\big|\int_{-\infty}^\infty dz\,\mathcal{E}_{0,z}(\Rb,z)\,\ee^{-\ii \omega_0 z/v}\big|$
determines the number of photons exchanged between the electron and the optical mode and can be evaluated through standard methods employed to compute EELS probabilities \cite{paper433}. We remark that, $\hat{\mathcal{S}}$ connects the density matrix in the interaction picture prior the scattering $\rho(-\infty)$ with the state after the interaction $\rho(\infty)$ as $\rho(\infty)=\hS\rho(-\infty)\hS^\dagger$. The operator $\hat{\chi}$ accounts for the non-resonant part of the electron-electron interaction mediated by the surrounding dielectric environment and induces an elastic phase shift on the wave function of a single electron passing close to a conductive surface \cite{paper357}. Owing to its short time scale in the few-fs range and the typical temporal separation between electrons of hundreds of fs, we disregard its effect in the rest of this work.  

The single-mode assumption underlying the validity of Eq. (\ref{scatt}) strongly depends on the value of $\beta_0$ for the coupling to each mode allowed by the material and the configuration details of the photonic structure collecting the electron emission. Generally, narrow-band selectivity can be achieved in one-dimensional geometries through phase-matching, when the mode's phase velocity $\omega_0/k_0$ equals the electron group velocity $v$, i.e., when $\omega_0/k_0\sim v$ \cite{KLS20,FHA22,AHT23,AHF24}. However, somewhat weaker selectivity can also be achieved in confined resonances supported by nanostructures \cite{ALM23,paper432}.  

To compute the statistical properties of the light emitted by electrons measured in a final set of longitudinal momenta $\qb_N=(q_1,\dots,q_N)$, we begin with the calculation of the matrix $T^{\qb_N}=\langle \qb_N |\, \hU \, \rho(-\infty)\,\hU^\dagger | \qb_N \rangle$, which is a key intermediate in the derivation of the optical density matrix. Indeed, it projects the evolved quantum state of the system (after interaction) onto the electron momentum eigenstates. Interestingly, its evaluation becomes straightforward when performed in the spatial representation $|\zb_N\rangle =\sum_{\qb_N}(\ee^{-\ii \qb_N\cdot \zb_N}/L^{N/2})|\qb_N\rangle$ (where $L$ is the quantization length), as these states satisfy the eigenequations $\hb|\zb_N\rangle =j(\zb_N)|\zb_N\rangle$ and $\hb^\dagger |\zb_N\rangle =j^*(\zb_N)|\zb_N\rangle$ with $j(\zb_N)=\sum_{i=1}^N \ee^{-\ii \omega_0 z_i /v}$. In physical terms, $j^*(\zb_N)$ represents the $\omega_0$-frequency component of a classical current in units of $-e$ formed by $N$ electrons longitudinally distributed as the components of $\zb_N$. As such, it is an eigenvalue of the current operator of negative frequency which is proportional to $\hb^\dagger$ \cite{paper373}. 

Under typical experimental conditions, the photonic structure is either in the vacuum state or excited with a laser, while the $N$-electron bunch exists in a complex state arising from an incoherent ensemble average over stochastic fluctuations of the electron source, combined with the coherent operations of IELS modulation and energy pre-filtering. To best describe such initial conditions, we set as pre-interaction electron-light state $\langle \zb_N|\rho(-\infty)|\zb_N'\rangle =\rho_e(\zb_N,\zb_N')\,|\alpha \rangle \langle \alpha |$, where $|\alpha\rangle$ is a bosonic coherent state of the mode with amplitude $\alpha$, and $\rho_e(\zb_N,\zb_N')$ is the spatial representation of the $N$-electron density matrix.

In order to account for general multi-electron post-filtering performed over a finite set of final momenta, we introduce the dimensionless detector function $F(\qb_N)$ which vanishes for values of $\qb_N$ outside the selected region. Thus, by integrating the product $F(\qb_N) T^{\qb_N}$, we can write the exact form of the output light density matrix after the interaction (see Appendix \ref{secA} for a detailed calculation):
\begin{align}
\rho_p=\frac{1}{P_F} \int d\zb_N d\zb_N' &\, \mathcal{F}(\zb_N-\zb_N')\rho_e(\zb_N,\zb_N')\label{ls1}\\
&\times \big|\alpha +\beta_0 j(\zb_N)\big\rangle \big\langle \alpha +\beta_0 j(\zb_N')\big|\nonumber,
\end{align}
where the function $\mathcal{F}(\zb_N)=\int d\qb_N F(\qb_N) \,\ee^{-\ii \qb_N\cdot \zb_N}/(2\pi)^N$ represents the detector response function. The normalization constant $P_F\leq 1$ corresponds to the probability of success of the post-filtering operation as well as the $N$-electron energy correlations developed during the light-mediated coupling \cite{K19,KLR24}. Importantly, Eq. (\ref{ls1}) establishes a direct connection between a generic incoming $N$-electron state and the created light state. Interestingly, the final optical density matrix is formed by a continuous superposition of coherent states with amplitudes determined by classical multi-electron currents and coefficients determined by the incoming $N$-electron state and the detector response function. Furthermore, Eq. (\ref{ls1}) highlights the possibility that a complete tomography of $\rho_p$ could enable full readout of $\rho_e(\zb_N,\zb_N')$, including the retrieval of quantum entanglement between the momentum states of different electrons. Entanglement has also been predicted to cause visible variations in the cathodoluminescence emission pattern when no post-filtering is applied \cite{KRA21_3}. 

Note that, if no post-filtering is performed [$\mathcal{F}(\zb_N-\zb_N')=\delta(\zb_N-\zb_N')$], Eq. (\ref{ls1}) shows that $\rho_p$ becomes  a function of the $N$-electron density $\rho_e(\zb_N,\zb_N)$ only. In this regime, the evaluation of expectation values of the normally-ordered light operators is made particularly simple. For high electron currents, Coulomb interaction through the propagation in TEM can induce marked  transversal and longitudinal energy correlations between electrons, as shown by a recent experiment measuring the ensemble properties of few-electron bunches \cite{HFD23}. While Eq. (\ref{ls1}) maintains its validity in the presence of such correlations, for illustrative purposes and to derive example results, in the following we consider the case of uncorrelated electrons, which is the case for sufficiently spaced electrons in time. In this scenario, the total density factorizes as $\rho_e(\zb_N,\zb_N)=\prod_{i=1}^N\rho^i_e(z_i,z_i)$, and all light properties depend on the so-called electron coherence factor (CF) \cite{paper371,paper374}
\begin{align}
M_k^i=\int_{-\infty}^\infty \!\!\! dz\, \rho^i_e (z,z)\,\ee^{\ii k z}.\label{cf}
\end{align}

\begin{figure}
\includegraphics[scale=1.]{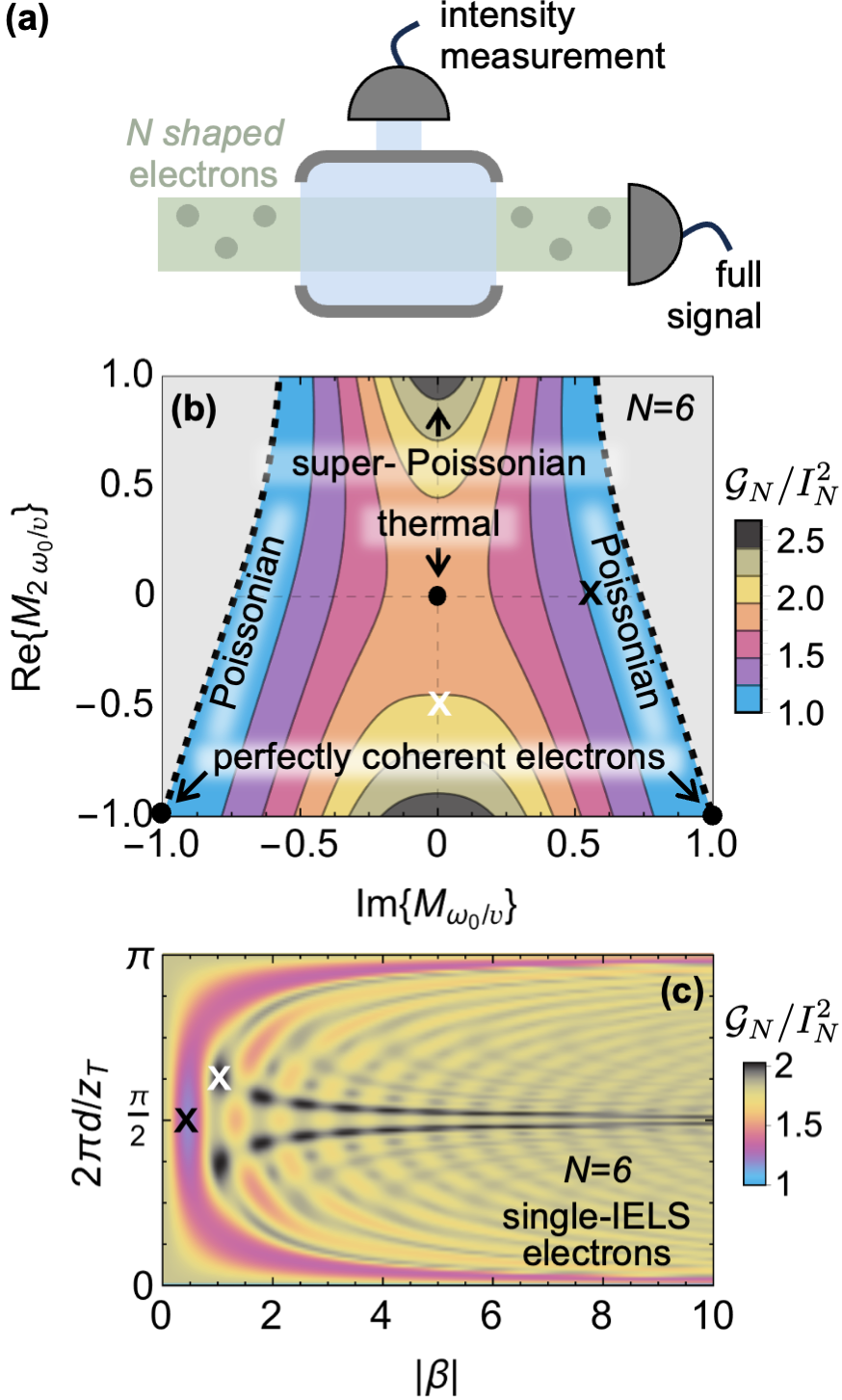}
\caption{{\bf Intensity fluctuations for $N$ modulated electrons}. Factor $\mathcal{G}_N/I_N^2$ defining the statistics of the emitted light without post-filtering [see Eq. (\ref{fluctu}) and sketch in panel {\bf (a)}] computed for $N=6$ electrons {\bf (b)}. The electrons are assumed to undergo the same modulation yielding coherence factor (CF) with imaginary $M_{\omega_0/v}$ and real $M_{2\omega_0/v}$ similarly to the CF after a IELS interaction [see Eq. (\ref{mfact})]. The grey areas correspond to unphysical electron states and CF values leading to negative light intensity fluctuations. {\bf  (c)} Same as in (b) but for electrons emitting light after a IELS modulation of strength $|\beta|$ and a subsequent free propagation of $d$ respect to the Talbot distance $z_{\rm T}=4\pi \me v^3\gamma^3/\hbar \omega_{\rm L}^2$  [see Eq. (\ref{mfact})]. The black and white crosses indicate the corresponding values of the CF in panel (b).}
\label{Fig5}
\end{figure}

The CF is a measure of the coherence carried by each of the electrons at momentum $k$, quantified through the strength of the Fourier components of their densities. In practice, it defines the ability of the light emitted by the electrons to interfere with a second time-varying signal \cite{paper374,paper373}. For instance, if all electrons share the same density ($M_k^i\equiv M_k$), the total radiated intensity in the absence of laser excitation takes the form $I_N=\langle \hn \rangle =\beta_0^2N\big[1+(N-1)|M_{\omega_0/v}|^2\big]$ (with $\hn=\ha^\dagger \ha$) and scales as $N^2$ when the CF approaches unity for all electrons. This multi-electron cooperative effect, where the interfering fields are mutually generated by the electrons, produces an emission intensity $\propto N^2$, resembling the Schwartz-Hora effect \cite{SH1969}, and is referred to as superradiance \cite{G05,GDL05}. Such behaviour has been experimentally observed in transition radiation \cite{LGF03} and is at the heart of radiation produced by free-electron lasers \cite{BC1985,BC1985_2,FA18}. The type of emission is also characterized by its intensity fluctuations $\Delta I^2_N=\langle \hn^2\rangle-I_N^2$ that read
\begin{align}
\Delta I_N^2/I_N=1+I_N\big[\mathcal{G}_N/I^2_N-1\big], \label{fluctu}
\end{align}
where $\mathcal{G}_N=\langle \ha^{\dagger 2}\ha^2\rangle$ is a function of only $M^i_{\omega_0/v}$ and $M^i_{2\omega_0/v}$ (see Appendix \ref{secB} for its exact form). Interestingly, it can only show super-Poissonian emission as the e-beam density is modified to give $\mathcal{G}_N/I_N^2>1$. Indeed, this conclusion can be drawn from the positivity of the fluctuations and the fact that, if $\mathcal{G}_N/I_N^2<1$, $\Delta I_N^2$ can assume an arbitrary negative value for strong enough coupling $\beta_0$, as $\mathcal{G}_N/I_N^2$ is independent of its value. 

In Fig. (\ref{Fig5}), we compute such important factor for $N=6$ identically modulated electrons yielding equal CF. In particular, we take electron densities leading to a pure imaginary and real CF at $k=\omega_0/v$ and $2\omega_0/v$, respectively. We motivate this particular choice after inspecting the form of the CF given by an electron after a single IELS modulation at $\omega_{\rm L}=\omega_0$ and a macroscopic propagation $d$ from the interaction zone
\begin{align}
M_{m\omega_0/v}=\ii^m {\rm sign}&\{\sin(2\pi m d/z_{\rm T})|\}^m \label{mfact}\\
&\times J_m[4|\beta\sin(2\pi m d/z_{\rm T})|]\nonumber
\end{align}
which can be calculated from the energy coefficients $c_\ell=J_\ell(2|\beta|)\ee^{\ii \ell {\rm arg}\{-\beta\}-2\pi \ii \ell^2 d/z_{\rm T}}$ using an envelope $\psi_0(z)$ spanning several optical cycles \cite{paper402,ZSF21,YFR21,paper374}. The $J_\ell(x)$ is the $\ell$-th Bessel function, $z_{\rm T}=4\pi \me v^3 \gamma^3 /\hbar \omega_{\rm L}^3$ is the Talbot distance, and $\beta$ is a coupling parameter analogous to $\beta_0$ but incorporating the electric field produced by the laser scattering off a material boundary \cite{PLZ10,paper360,paper371}. We observe a wide range of possible CF leading to $\Delta I_N^2/I_N \approx 1$ for all $N$ explored [see Fig. (\ref{Fig5}b) for $N=6$], even though a higher number of electrons would lead to larger departure from the Poissonian regime given the stronger intensities. For electrons with vanishing CF, we observe $\Delta I_N^2/I_N\approx 1+I_N$, typical signature of a thermal light-emitting source \cite{BC1985,ZIF24}. Interestingly, all these light statistics can be harnessed through careful choice of the IELS parameters, as shown in Fig. (\ref{Fig5}c).

\begin{figure*}
\includegraphics[scale=0.72]{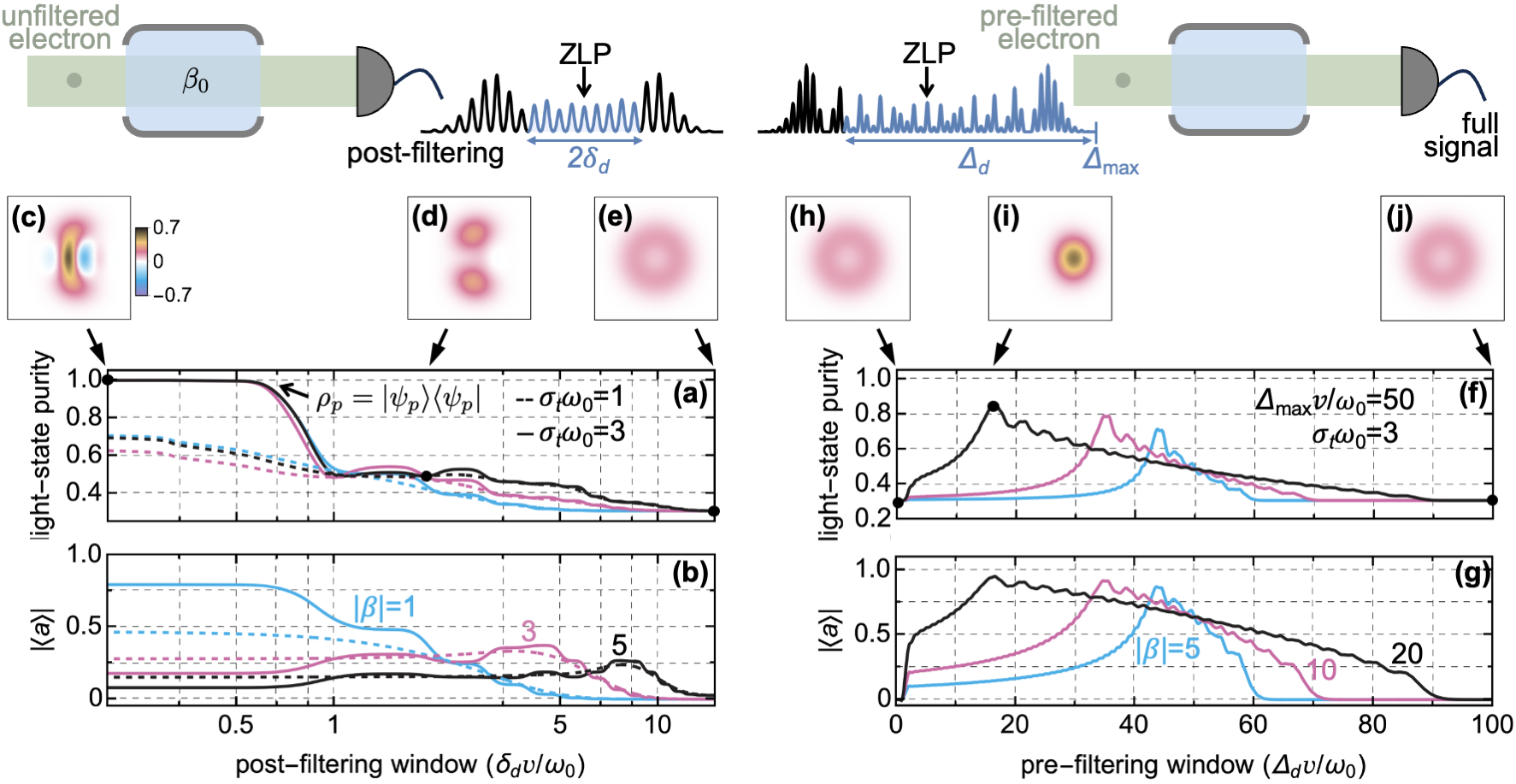}
\caption{{\bf Properties of the light state generated by single electrons using energy post- or pre-filtering}. An electron with an incoherent envelope of temporal width $\Delta t \omega_0 \gg 1$ travels a negligible distance $z_{\rm T}=4\pi \me v^3\gamma^3/\hbar \omega_{\rm L}^2$, i.e., $d/z_T=0$, from a single IELS interaction of strength $\beta$ and frequency $\omega_{\rm L}=\omega_0$ to couple with an optical mode with strength $\beta_0=1$ initially in a vacuum state $|0\rangle \langle 0|$. After the interaction, the light state purity {\bf (a)} and the absolute value of the average of the photonic creation operator {\bf (b)} are computed by considering the electrons with normalized coherence time $\sigma_t \omega_0=1$ (dashed lines) and $3$ (solid lines) and longitudinal momentum in a window $2\delta_d$ symmetric around the zero-loss peak (ZLP), as shown in the post-sample asymmetric spectrum above panels (c-e). {\bf (c,d,e)} Photonic Wigner function after coupling with an electron with $\sigma_t\omega_0=3$ for the post-filtering windows $\Delta_d v/\omega_0=0.01$, 2, 15, respectively. {\bf (f,g)} Same as (a,b) with $\sigma_t \omega_0 =3$ but discarding the electrons outside the momentum range between $\Delta_{\rm max}-\Delta_d$ and $\Delta_{\rm max}=50\omega_0 /v$ immediately after a IELS stage, as shown in the symmetric spectrum above panels (h-j), and without final energy post-filtering. {\bf (h,i,j)} Photonic Wigner function corresponding to the pre-filtering windows at $\Delta_d v/\omega_0=$0.01, 16.5, 100, respectively. In all panels, we use ${\rm arg}\{-\beta\}=0$.}
\label{Fig2}
\end{figure*}

A more complex situation is found for a general post-sample filtering function. In this case, the number representation $\rho_p=\sum_{nn'} \rho_{p,nn'}|n\rangle\langle n' |$ provides a clearer isolation of the role played by the input electron density matrix, which is otherwise obscured in the spatial dependence of the coherent states in Eq. (\ref{ls1}). While again considering uncorrelated electrons and an initial vacuum state ($\alpha=0$), we calculate $\rho_{p,nn'}$ from Eq. (\ref{ls1}) through a combinatorial analysis leading to  the following result (see Appendix \ref{secB} for a detailed derivation)
\begin{align}
\rho_{p,nn'}=&\frac{1}{P_F} \!\!\!\sum_{\substack{k,k',\mb\\ \mb',\pb,\pb' \geq 0}}\mathcal{C}_{\mb,\mb',\pb,\pb'}^{(n,k,n',k')}\int d\qb_N F(\qb_N)\label{rhonnpN} \\
&\times   \prod_{i=1}^N \PM^i_{\omega_0(m_i-m_i')/v}\Big[q_i+\frac{\omega_0}{2v}(m_i+m_i')\Big],\nonumber
\end{align}
where the $\beta_0$-dependent coefficients $\mathcal{C}_{\mb,\mb',\pb,\pb'}^{(n,k,n',k')}$ are defined in Eq. (\ref{combc}) and their specific form is not of fundamental relevance to this work. The vectors $\mb,\mb',\pb$ and $\pb'$ are composed by positive integers and have dimension $N$. Interestingly, Eq. (\ref{rhonnpN}) condenses the electron dependence into the factor 
\begin{align}
\PM^i_{k}(q)= \int_{-\infty}^\infty \!\! \!\! dz \,W^i_e(z,q)\,\ee^{\ii k z}\label{pcf}
\end{align}
which we term projected coherence factor (PCF), as it plays a role similar to the CF when only a sub-set of scattering events are observed and it is defined through the electron Wigner function $W^i_e(z,q)=\int_{-\infty}^\infty \!dy\, \rho^i_e(z-y/2,z+y/2)\,\ee^{\ii q y}/2\pi$ \cite{W1932} representing the quantum analogue of a classical phase-space density. Equation (\ref{pcf}) reveals that when final energies are measured, the electron density involved in the interaction is only determined {\it a posteriori} through the post-filtering procedure. Specifically, the spatial frequencies that influence $\rho_{p,nn'}$ are those arising from the Fourier transform along the propagation axis of the density obtained through the integration of the electron Wigner function over the finite momentum range set by $F(\qb_N)$. In Fig. (\ref{FigS1}a), we illustrate the sub-cycle structuring of several such cuts of the Wigner function corresponding to an IELS-modulated electron, also measured through a reconstruction algorithm based on a double-IELS interaction scheme \cite{PRY17}. Reassuringly, when no post-filtering is applied, the momentum integral of the PCF coincides with the CF, namely, $M^i_{k}=\int_{-\infty}^\infty\! dq \,\PM^i_{k}(q)$, as is directly evident from the Wigner function definition.  

\subsection{Light-state purity and electron coherence \label{secb}}
An ideal quantum state, unaffected by classical ensemble averages over initial conditions or mechanisms of decoherence, can be described by a pure state $|\psi_p\rangle=\sum_{n=0}^\infty \alpha_{p,n}|n\rangle$ and, equivalently, by the density matrix $\rho_p=|\psi_p\rangle \langle \psi_p|$. Here, we aim to explore how electron coherence and post-filtering determine the final purity of the light.

First, we examine Eq. (\ref{ls1}) in the case of uncorrelated electrons (although this assumption is not necessary for the following statement to hold) and observe that, if an infinitely precise post-filtering measurement with outcome $\tilde{\qb}_{N}$, described by $F(\qb_N)\sim\delta(\qb_N-\tilde{\qb}_{N})$, is performed, $\rho_p$ becomes perfectly pure, provided the electron state is also pure, i.e., $\rho^i_e(z_i,z_i')=\psi^i_e(z_i)\psi^{i*}_e(z_i')$. In most experiments performed in SEM/TEM, the latter assumption is not met because electrons arrive at the sample at a time $t_{0,i}$ that can incoherently fluctuate by $\Delta t \sim 100$ fs \cite{PKZ12,KGK14,FBR17,HDR24}. However, since they have coherence times $\sigma_t\sim 5$ fs spanning several optical cycles ($\sigma_t \omega_0\gg1$), their PCF is not affected by the incoherent averaging at the spatial frequencies of interest for this work $k=m\omega_0/v$, with $m$ an integer number, therefore effectively providing the aforementioned purity condition (see Appendix \ref{secB} for a detailed proof). Thus, we can conclude that, regardless of the specific form of the coherently modulated electron state, the determination of the final energies of all electrons guarantees a pure light state. However, such purity will be maintained over the spectral width $\sim \hbar /\Delta t\sim 10$ meV around $\omega_0$.

We now examine this result in the simple case of a single electron, for which Eq. (\ref{rhonnpN}) simplifies to the form (see Appendix \ref{secC})
\begin{align}
& \rho_{p,nn'} =\frac{1}{P_F}  \langle n|\beta_0\rangle \langle \beta_0|n'\rangle \label{state1el} \\
&\quad \times \int_{-\infty}^\infty \!\!\!\! dq\,F(q) \,\PM_{\omega_0 (n'-n)/v}[q+\omega_0(n+n')/2v].\nonumber 
\end{align}
In Fig. (\ref{Fig2}a), we analyze the purity ${\rm Tr}\{\rho_p^2\}$ of the state in Eq. (\ref{state1el}) for an electron with coherent Gaussian envelope of standard deviation $\sigma_t$ and incoherent ensemble distribution of width $\Delta t \omega_0 \gg 1$ modulated through an IELS stage of laser frequency $\omega_{\rm L}=\omega_0$, and subsequently propagated over a distance $d$ from the interaction zone, as done to obtain Eq. (\ref{mfact}). As expected, the light-state purity approaches unity when the post-filtering window $2\delta_d$, collected by the energy detector, is $\delta_d \sigma_t v \lesssim 1$ as long as the electron coherence spans several optical cycles, while it stabilizes to the fully-mixed value $\sum_{n=0}^\infty \rho_{n n}^2$, when the post-filtering window covers the entire electron spectrum. This result is in agreement with the form of the $m$-th order CF in Eq. (\ref{mfact}), vanishing for $d/z_{\rm T} \sim 0$ and $m\neq 0$, and the generated light state $\rho_{p,nn'}=  \langle n|\beta_0\rangle \langle \beta_0|n'\rangle M_{\omega_0(n'-n)/v}$ obtained from Eq. (\ref{state1el}) in the $\delta_d \rightarrow \infty$ limit. Accordingly, the form of the photonic Wigner function \cite{MW95}, also showing negative values, represent a pure quantum state generated by a IELS electron for small $\delta_d$ and a phase-averaged coherent state where the entire spectrum is considered [see Fig. (\ref{Fig2}a-e)].

\begin{figure}
\includegraphics[scale=1.]{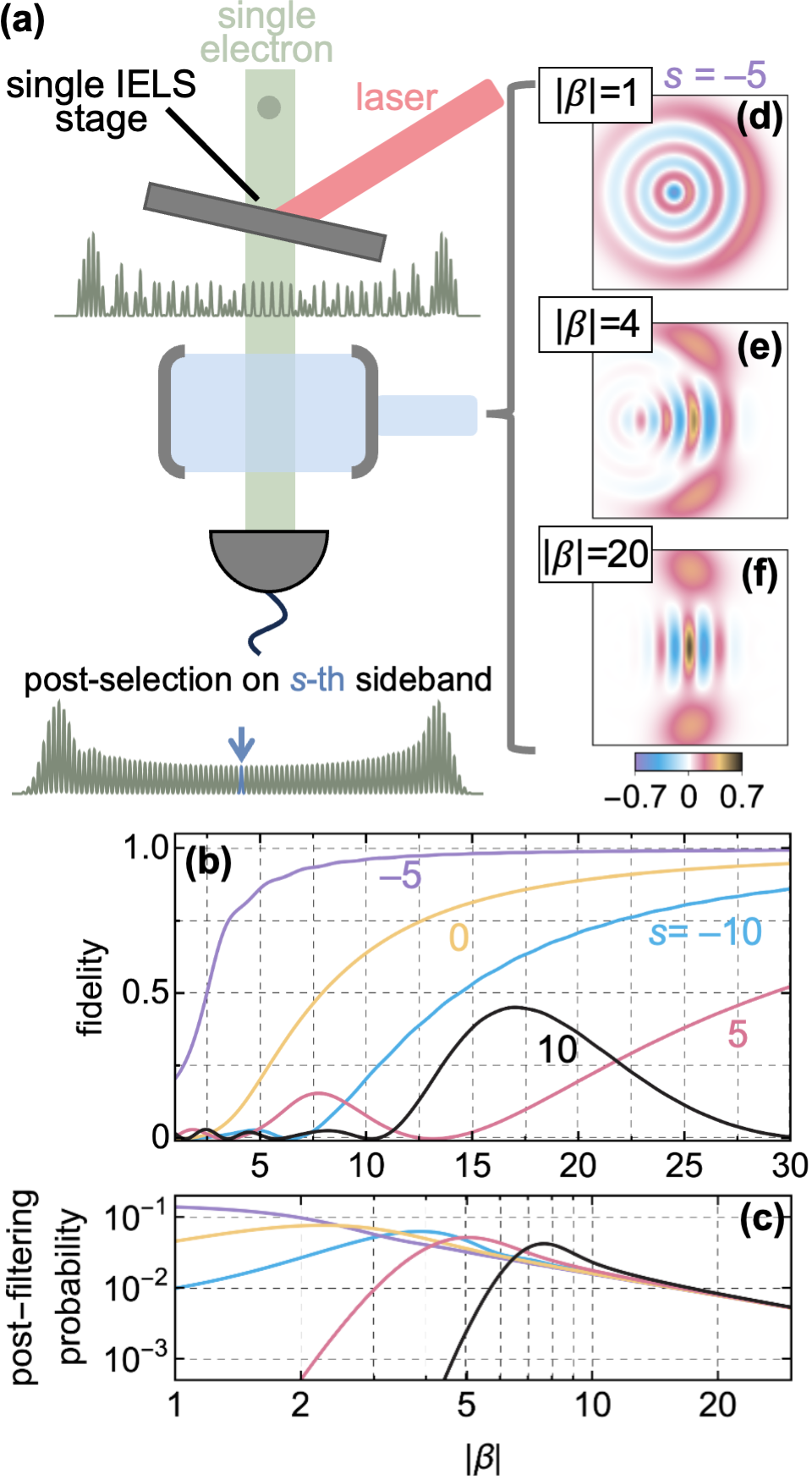}
\caption{{\bf Natural formation of cat states after a single IELS stage}. {\bf (a)} Proposed scheme to produce high-purity cat states from an optical mode in a vacuum state involving a single IELS interaction of coupling parameter $\beta$ and the post-filtering of the $s$-th sideband after spontaneous emission into the cavity with strength $\beta_0=2$. {\bf (b)} Overlap between the light state generated by an electron after passing through the stages sketched in (a)  and a cat state with $\chi=-\ii \beta_0\ee^{\ii \,{\rm arg}\{-\beta\}}$, $\theta=s\pi+\pi/2-4|\beta|$ for different IELS couplings $\beta$ and post-filtered sideband order $s$. {\bf (c)} Post-filtering probabilities for the configurations reported in (b). {\bf (d-f)} After-interaction photonic Wigner function for $s=-5$ and $|\beta|=1$, 4, 20. In all panels, we use ${\rm arg}\{-\beta\}=0$.}
\label{Fig3}
\end{figure}

As we previously observed, in addition to enabling access to high-purity states, the combination of post-filtering and shaped electrons provides a means to probe time-varying signals with an electron density that depends on its final measured energy and that can be visualized through the energy cuts of the electron Wigner function [see Fig. (\ref{FigS1}a)]. An example of this is the average electric field $\langle \hat{\Eb} (\rb) \rangle=\vec{\mathcal{E}}_0(\rb)\langle \ha \rangle +\vec{\mathcal{E}}^*_0(\rb)\langle \ha^\dagger  \rangle \propto |\langle \ha \rangle |$ emitted by the electron into the light mode, which varies as a function of $\delta_d$ [see Fig. (\ref{Fig2}b)]. This capability could be particularly significant for studying and controlling the dynamics in materials \cite{CZ09,SF15_2} triggered by the same laser used to modulate the beam with sub-ps precision.

A similar phenomenon of enhanced time localization occurs when an energy filter, selecting a fixed momentum range starting from $\Delta_{\rm min}=\Delta_{\rm max} -\Delta_d$ and ending at $\Delta_{\rm max}$ relative to the central momentum, is placed between the IELS modulation and the interaction with the sample [see Fig. (\ref{Fig1}) and the rightmost sketch in Fig. (\ref{Fig2})]. Indeed, since the CF can be re-expressed in terms of the PCF of an electron without  pre-filtering $\PM^{\rm unf}_k$ as  
\begin{align}
M_k = \frac{1}{M_0}\int^{\tilde{\Delta}_{\rm max}}_{\tilde{\Delta}_{\rm min}} \!\!\!\!\!\!\!dq \,\PM^{\rm unf}_{k}(q+k/2)\label{mkunf}
\end{align}
with $\tilde{\Delta}_{\rm max}={\rm min}\{\Delta_{\rm max},\Delta_{\rm max}+k\}$ and $\tilde{\Delta}_{\rm min}={\rm max}\{\Delta_{\rm min},\Delta_{\rm min}+k\}$, this procedure effectively corresponds to selecting an energy portion of $W_e(z,q)$, thereby influencing the involved electron density and its related quantities, such as the average electric field [see Fig. (\ref{Fig2}g)]. The factor $M_0$ represents the probability of pre-filtering and guarantees wave function normalization. The resulting enhanced electron coherence is also reflected by the light-state purity depicted in Fig. (\ref{Fig2}f) for an electron pre-filtered right after ($d=0$) a IELS interaction. There, we observe several maxima (with $\sim 0.86$ the greatest value), each one for a given energy window  $\hbar \Delta_d v$ and coupling strength $\beta$ as well as a convergence to the mixed-state value for small and large $\Delta_d$. This behavior can be understood by examining the corresponding CF in the $\sigma_t \omega_0 \gg 1$ limit, expressed as [see Eq. (\ref{filtM})]:
\begin{align}
|M_{m\omega_0/v}|&=\label{CFfilt} \\
\!\Bigg| &\sum_{\ell=\ell_{\rm min}}^{\ell_{\rm max}}\!\! J_\ell(2|\beta|)J_{\ell+m}(2|\beta|)\,\ee^{4\pi \ii m\ell d/z_{\rm T}}/M_0\Bigg|,\nonumber
\end{align}
where $\ell_{\rm min}=\lfloor (\Delta_{\rm max}-\Delta_d)v/\omega_0 \rfloor-\min\{0,m\}$ and $\ell_{\rm max}=\lfloor \Delta_{\rm max}v/\omega_0 \rfloor-\max\{0,m\}$, and $\lfloor x\rfloor$ denotes the integer part of $x$. This expression reveals a significant increase in electron coherence, surpassing the absolute maximum of $|M_{\omega_0/v}|\sim 0.58$ observed in bunched densities following an IELS interaction and a drift in free space \cite{PRY17,B17_2,paper360}. For instance, with $|\beta|\sim 20$, we achieve $|M_{\omega_0/v}|\sim 0.95$ for various values of $d$, including $d/z_{\rm T}\sim0$ [see Fig. (\ref{FigS1}b-d)]. Given the macroscopic lengths on the centimeter scale required by standard energy filters to operate, such case refers to an idealized scenario not achievable experimentally in a straightforward manner. However, at Talbot revivals and thus larger distances, depending on the coherence time and IELS strength, similar results could be achieved. In particular, optimal purity is achieved  by filtering near the lobes of the IELS energy distribution, as in that region the electron density confines to a limited range in time [see Fig. (\ref{FigS1}a)]. Importantly, this type of strategy can also be used as an alternative approach to pulse compression \cite{PRY17,KSH18}.

Despite this high coherence for low $m$, Eq. (\ref{CFfilt}) vanishes for $\lfloor \Delta_dv/\omega_0 \rfloor \leq |m|$, thereby limiting the light-state purity in a manner dependent on the electron-mode coupling $\beta_0$. Finally, as previously demonstrated \cite{paper360}, $\rho_{p}$ oscillates between a quasi-pure and a phase-averaged coherent state as electron coherence is varied through $\Delta_d$ [see Fig. (\ref{Fig2}h-j)].

Finally as expected, for nearly elastic attosecond imaging or diffraction experiments, it also becomes irrelevant if the filtering takes place before or after the sample. This is confirmed by the $k\rightarrow 0$ limit of the integral in Eq. (\ref{mkunf}) that transforms to an integrated PCF over the collection range as it appears in Eqs. (\ref{rhonnpN},\ref{state1el}) for negligible $\omega_0$. 

\subsection{Natural synthetization of cat states by IELS electrons \label{secc}}
We now utilize the purity achieved through post-filtering performed around the $s$-th energy sideband in the high electron coherence limit of Fig. (\ref{Fig2}a), to examine the actual state of the generated light [see Fig. (\ref{Fig3}a)]. Under these conditions, Eq. (\ref{state1el}) predicts that any form of electron energy shaping will yield $\rho_p=|\psi_p\rangle\langle \psi_p|$ with expansion coefficients in number basis directly connected to the $c_\ell$ as     
\begin{align}
\alpha_{p,n}=\frac{\langle n|\beta_0\rangle\, c_{n+s}}{\sqrt{\sum_{n=0}^\infty |\langle n|\beta_0\rangle \,c_{n+s}|^2 }}.\label{alphan}
\end{align}
Equation (\ref{alphan}) demonstrates that any target light state with finite support can be synthesized through appropriate shaping of the electron energy coefficients $c_{\ell}$. Intuitively, it predicts an average photon number that depends on $\beta_0$ but can exceed the probability of spontaneous emission, $\beta_0^2$. This effect arises from the post-filtering process, where only a subset of events is considered during the photon measurements, and is related to the weak value of a quantum observable \cite{AAV1988}.

In the special case of an electron immediately after a one-stage IELS interaction ($c_\ell=J_\ell(2|\beta|)\ee^{\ii \ell {\rm arg}\{-\beta\}}$), we find that, beyond a certain high value of $|\beta|$, the electron naturally forms an approximate version of a cat state, $\alpha_{p,n}\propto\langle n|\chi \rangle[1+\ee^{\ii \theta}(-1)^n]$, where $\chi=-\ii \beta_0\ee^{\ii {\rm arg}\{-\beta\}}$ and $\theta=s\pi+\pi/2-4|\beta|$. Taking this state as the target state $|\psi_p^{{\rm targ}}\rangle=\sum_{n=0}^\infty\alpha_{p,n}^{\rm targ}|n\rangle$, we compute its overlap with $|\psi_p\rangle$ using the fidelity $|\langle \psi_p |\psi_p^{{\rm targ}}\rangle|^2$. Remarkably, this shows near-perfect generation under the condition $(n_{\rm max}+s)^2/2\ll |\beta|$, determined by the first $n_{\rm max}$ coefficients required to accurately describe $|\psi_p^{{\rm targ}}\rangle$, which is itself set by the value of $|\chi|=\beta_0$ [see Fig. (\ref{Fig3}b)]. Such high coupling strengths have already been experimentally demonstrated with pulsed-laser interactions near a nanostructure \cite{DNS20} and in free space \cite{KES18} as well as under continuous-wave seeding of a Si${}_3$N${}_4$
microresonator \cite{HRF21}. However, due to the large energy spread introduced by the $|\beta|\gtrsim 10$ IELS interaction, post-filtering probabilities are found to be $\lesssim 1\%$ [see Fig. (\ref{Fig3}c)] at fidelities $\gtrsim 99\%$ [see Fig. (\ref{Fig3}d-f)]. 

\begin{figure*}
\includegraphics[scale=0.66]{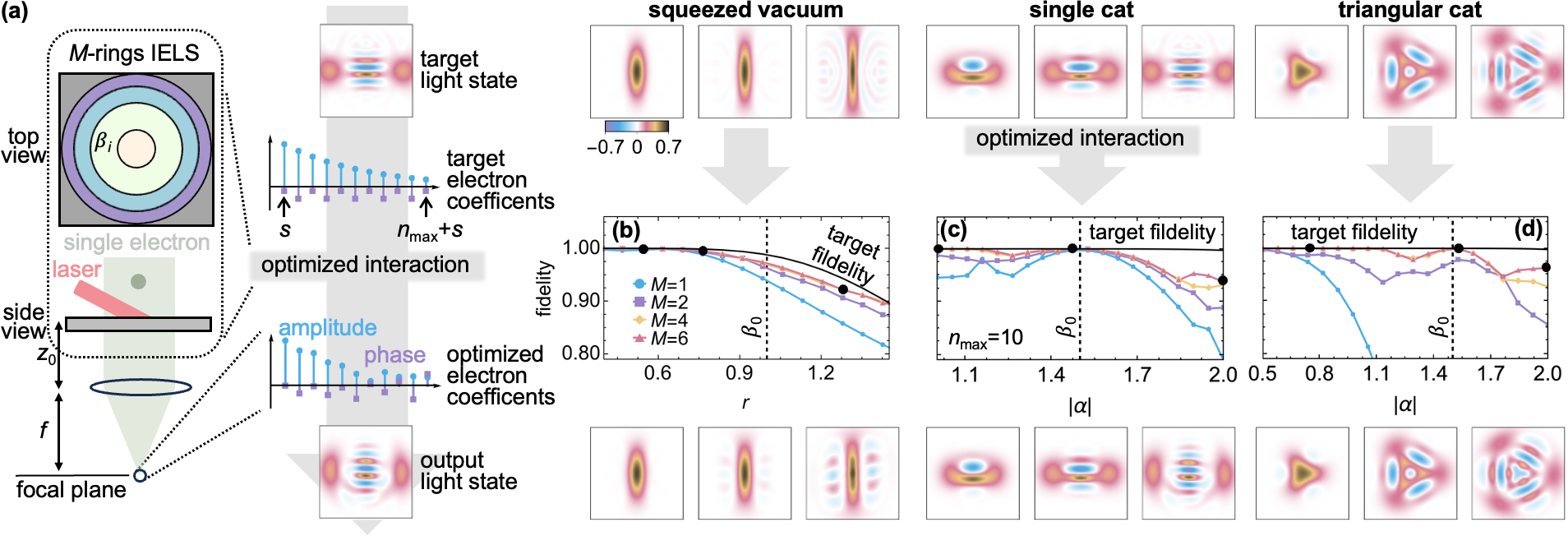}
\caption{{\bf Lateral IELS patterning for electron energy coefficients optimization and quantum light generation}. {\bf (a)}. Illustration of the steps employed for tailored synthesis of quantum light states. A set of electron energy amplitudes are obtained from Eq. (\ref{alphan}) to approximate the first $n_{\rm max}=10$ coefficients of a given target photonic state $|\psi^{{\rm targ}}_p\rangle$ and employed to optimize the design of the radial profile of the near-field used in an IELS stage composed of $M$ concentric rings each one with constant $\beta_i$. The most favorable design is supposed to provide the electron energy coefficients producing the optimal light state $|\psi^{{\rm opt}}_p\rangle$ that maximizes the fidelity $|\langle \psi^{{\rm targ}}_p|\psi^{{\rm opt}}_p\rangle|^2$. {\bf (b-d)} Maximum achieved fidelity for $M=1,2,4,6$ concentric rings for different types of light state: a squeezed vacuum with coefficient $r$ (a), a cat state with real amplitude $\alpha$ and phase $\theta=\pi/2$ (b), and a triangular cat state with real amplitude $\alpha$ and $\theta=2\pi/3$ (c). The light Wigner functions on the top row correspond to target states while the ones in the bottom to generated states in the configurations highlighted by the black circles in (b,c,d). A laser modulation frequency of $\omega_{\rm L}=2\omega_0$ was used for (b) and of $\omega_{\rm L}=\omega_0$ for (c,d) as well as $\beta_0=1$ and $1.5$, respectively.}
\label{Fig4}
\end{figure*}

\subsection{On-demand quantum light generation by lateral IELS \label{secd}}
The approach previously used to create a specific type of cat state can be generalized to a broader range of light states through Eq. (\ref{alphan}) by accessing a wider set of electron energy coefficients $c_\ell$. Several schemes have been proposed to achieve such flexibility, primarily relying on either sequential combinations of IELS and free propagation stages \cite{YFR21} or focusing different lateral sections of an e-beam that has passed through a spatially dependent coupling coefficient $\beta(\Rb)$ \cite{paper415}. In this work, we adopt the latter approach whose capabilities are reported in Fig. (\ref{Fig4}); however, a similar study could be conducted following the former method.

As detailed in Appendix \ref{secD}, the energy coefficients forming the wave function near the focal point of a lens acting on an electron previously shaped by a near field divided into $M$ equal-area circular sectors, each producing constant IELS coefficients $\beta_i$ [see the modulation scheme in Fig. (\ref{Fig4}a)], are given by
\begin{align}
c_\ell=\ee^{-2\pi \ii\ell^2 d/z_{\rm T}}\sum_{i=1}^M J_\ell(2|\beta_i|)\ee^{\ii\ell{{\rm arg}\{-\beta_i\}}},\label{lIELS}
\end{align}
where now $d=z_0+f$ is the sum of the lens' separation from the IELS plane ($z_0$) and the focal distance ($f$). We use an optimization algorithm based on a steepest descent routine (see Appendix \ref{secD} for details) to determine the set of coupling strengths $\beta_i$, lens position $d$, and post-filtering sideband $s$ that maximize the overlap of the generated state $|\psi_p^{{\rm opt}}\rangle$ with a given target light state. This is achieved by repeatedly inserting Eq. (\ref{lIELS}) into Eq. (\ref{alphan}) [see Fig. (\ref{Fig4}a)]. Specifically, the optimization process runs over $c_{s},\dots,c_{n_{\rm max}+s}$ while verifying that the inclusion of additional coefficients does not result in any significant changes.

As target states, we select the first $n_{\rm max}$ coefficients, which define a maximum achievable target fidelity [black solid lines in Fig. (\ref{Fig4}b-d)], for a squeezed vacuum with $\alpha_{p,2n}^{\rm targ} \propto (-\tanh r)^{n}\sqrt{(2n)!}/2^n n!$, a cat state $\alpha_{p,n}^{\rm targ} \propto \langle n|\alpha\rangle[1+\ee^{\ii \theta}(-1)^n]$, and a triangular cat state with $\alpha^{\rm targ}_n \propto \langle n|\alpha\rangle[1+\ee^{\ii n\theta}+\ee^{2\ii n\theta}]$ [see the first row of photonic Wigner functions in Fig. (\ref{Fig4}b-d)]. However, we remark that this method is applicable to any set of coefficients $\alpha^{\rm targ}_{p,n}$. 

For the squeezed vacuum, we achieve fidelities of nearly $100\%$ for amounts of squeezing smaller than $\beta_0$ by modulating the electron at twice the fundamental frequency ($\omega_{\rm L}=2\omega_0$), which suppresses the emission of an odd number of photons for even $s$, simplifying optimization. While this result is largely independent of the number of sectors for small $r$, when the average number of required photons exceeds $\beta_0^2$, we observe a significant improvement in synthesizing the target state as $M$ increases [see Fig. (\ref{Fig4}b)].

For cat and triangular cat states, the ability of the coefficients in Eq. (\ref{lIELS}) to replicate $\alpha_{p,n}^{\rm targ}$ improves dramatically with the addition of more circular sectors, raising fidelity from below $80\%$ for $M=1$ to nearly $100\%$ for $M=6$ [see Fig. (\ref{Fig4}c,d)]. Within the explored parameter range, the optimal IELS couplings are confined to the range $0 \lesssim |\beta_i| \lesssim 14$ [in Fig. (\ref{FigS2}), we report their values], while post-filtering probabilities range from $10\%$ to $0.1\%$, depending on whether $\langle \psi_p^{\rm targ}|\hn |\psi_p^{\rm targ}\rangle$ is smaller or larger than $\beta_0^2$, respectively.

\section{Discussion and concluding remarks \label{secz}}
In this work, we have presented a compact theoretical framework that enables the study of the light state generated by the interaction of $N$ pre-modulated electrons with a single optical mode, within a specific subset of scattering events selected by a final electron spectrometer [see Fig. (\ref{Fig1})].

We have demonstrated that, without final energy filtering, the resulting light density matrix $\rho_p$ can exhibit super-Poissonian statistics due to inter-electron photon exchange, but its purity is strongly constrained by the electron coherence, quantified by the absolute value of the coherence factor (CF) $M^i_k$, i.e., the strength of the Fourier components of the single-electron density $\rho^i_e(z,z)$ [see Eq. (\ref{cf})]. To enhance the CF to approximately $95\%$, we proposed retaining only the electrons exiting a strong ($|\beta|\sim 20$) IELS modulation with energies inside a specific window, which effectively compresses the e-beam temporally. The advantage of this scheme, compared to others that combine longitudinal \cite{YFR21} or later IELS interactions \cite{paper415}, is that it relies only on a single homogeneous IELS stage -- a resource increasingly common in ultrafast TEM -- and an energy filter, such as a Wien filter \cite{TI13}, placed before the sample rather than after, as in energy-filtered EELS measurements \cite{VVV04}. At optical frequencies, the optimal energy window is approximately $20$ eV [see Fig. (\ref{Fig2}f)], making the filtering requirements less stringent than in such experiments. Using this practical scheme for a single electron, we have shown that coherent states with a purity of approximately $90$ \% can be generated [see Fig. (\ref{Fig2}g)].

We have also examined how $\rho_p$, and the associated light properties, are influenced by electron modulation when post-filtering is applied to a specific kinetic energy window. Specifically, we found that electron coherence is now quantified by the projected coherence factor (PCF) [see Eq. (\ref{pcf})], where the electron density appearing in the CF is replaced by the electron Wigner function $W^i_e(z,q)$ integrated over a specific range of momenta. Since this range is selected {\it a posteriori}, this result demonstrates how different post-filtering windows can reveal information about a specimen probed through various sub-cycle density modulations. In terms of light state purity, we demonstrated that for any electron modulation yielding the energy coefficients $c_\ell$, a narrow post-filtering window produces a perfectly separable state, even under stochastic electron illumination with random times of arrival, provided the electrons have coherence times spanning several optical periods [see Fig. (\ref{Fig2}a)].

By leveraging this result, we have then demonstrated several cases where quantum light can be harnessed using only a single IELS stage. We showed how cat states can be generated without lateral patterning of the IELS field or dispersive electron compression, achieving $\sim 100\%$ fidelity with probabilities exceeding $1\%$ [see Fig. (\ref{Fig3}b,c)]. Furthermore, to synthesize other types of light states, we proposed a scheme based on optimizing the $c_\ell$ coefficients produced by an IELS interaction composed of $M$ concentric sectors [see Eq. (\ref{lIELS})]. Applying this approach to the generation of squeezed vacuum, cat, and triangular cat states, we demonstrated that $M=6$ sectors are sufficient to achieve their production with $\sim 99\%$ fidelity and probabilities greater than $0.1\%$, provided the required average number of photons remains close to the Poissonian spontaneous emission average $\beta_0^2$.  

In all analyzed cases, the creation of light states with strong quantum features, such as high squeezing or Wigner function negativity, requires a high average photon number, which in turn necessitates above-unity values of $\beta_0$. Recent experiments with electrons passing extended structures of about $\sim 100$ $\mu$m in length reported photon generation in a dielectric waveguide at an average coupling parameter of $\beta_0\sim0.32$ \cite{AHF24}, and EELS at a hybrid metal-dielectric multilayer structure corresponding to $\sim 0.99$ \cite{AHT23}. Higher coupling strengths are expected for longer interaction lengths \cite{paper433}. However, in the current optimization scheme [Fig. (\ref{Fig4}a)], the electron coefficients maintain the form reported in Eq. (\ref{lIELS}) only over a distance of approximately $\lambda_e/{\rm NA}^2$, suggesting small numerical apertures at high energies such as NA$\sim 2\times 10^{-4}$ at $E_0^e=100$ keV. Alternatively at lower kinetic energies and for larger numerical apertures, infrared plasmonic resonances with dimensions $D$ similar to this scale, such as those found in nanostructured two-dimensional materials \cite{YLZ13,paper303,paper433}, may be preferred. Since $D\sim 1/k_0$, this conclusion is further supported by the phase-matching condition $\omega_0 /k_0 v\sim 1$, which suggests low electron velocities for small-sized structures. If this detrimental effect remain a problem, its consequences could be mitigated by explicitly accounting for it in the optimization process. 

Another possibility to increase the bare coupling strength $\beta_0$ is offered by the application of the optimization scheme to $N$-electron pulses, leveraging the superradiant enhancement to achieve an effective coupling strength of $\sim N\beta_0$. In practice, such implementation only requires the use of Eq. (\ref{rhonnpN}) in the $\sigma_t \omega_0 \gg 1$ and exact post-filtering limits [see Eq. (\ref{sealpha})], in order to compute the fidelity between target and emitted light states. The exploration of this approach is left for future work.

We believe the analysis presented here represents a significant step towards a more comprehensive understanding of $N$-electron emission into free space and photonic structures under strong coupling conditions and the practical realization of tunable sources of complex quantum light states in photonic devices.

\acknowledgments 
We thank Hao Jeng, Aviv Karnieli, and F.~Javier~Garc\'{\i}a~de~Abajo for helpful and enjoyable discussions. This work has been supported in part by the European project EBEAM (ID:101017720).

\section*{Author contributions}

V.D.G. and C.R. provided the ideas. V.D.G. developed the theory, performed the simulations, and produced the figures. All authors discussed the results and interpretation. V.D.G. wrote the manuscript, with contributions and input from R. H. and C.R.

\section*{Competing Interests}
The authors declare no competing interests.

\appendix
\section*{Appendix}

\section{Photonic density matrix after interaction with an $N$-electron pulse and successive post-sample energy filtering}

\label{secA}
\renewcommand{\theequation}{A\arabic{equation}}
\renewcommand{\thesubsection}{\arabic{subsection}}

In this section, we want to evaluate the density matrix associated with a single optical mode of frequency $\omega_0$ after the interaction with an electron beam composed by $N$ relativistic electrons, all with central kinetic energies $E^e_0\gg \hbar \omega_0$ corresponding to a velocity $\vb=v \zz=\hbar q_0 \zz /\gamma \me$, where $\gamma=1/\sqrt{1-v^2/c^2}$, and the action of a post-sample energy filtering (post-filtering) performed by an electron spectrometer. In what follows, we will assume the interaction to happen along the propagation direction of the electron bunch crossing the transverse position $\Rb$ at some instant of time. 

For the energies analyzed in this work, the electrons do not change considerably during the interaction time and as a consequence their dispersion relation $E_q=c\sqrt{\me^2 c^2 + \hbar^2 q^2}$ can be expanded to retain only the first linear term as $E_q\approx \me c^2+E^e_0 + \hbar\vb\cdot (\qb-\qb_0)$. Under this assumption, also known as nonrecoil approximation, the scattering operator $\hat{\mathcal{S}}(t,-\infty)$ associated with the system dynamics can be worked out explicitly \cite{paper373} and in second quantization it takes the form 
\begin{align}
\hat{\mathcal{S}}(\infty,-\infty)= \ee^{\ii \hat{\chi}}\,\ee^{\int_0^\infty \! d\omega\, g_\omega (\hb_\omega^\dagger \ha_\omega-\hb_\omega \ha_\omega^\dagger)}. \label{scattSI}
\end{align}
In Eq. (\ref{scattSI}) $\hat{\chi}$ is an operator that accounts for nonresonant electron-electron coupling mediated by the electromagnetic environment and $\hb_\omega=\sum_k \hc^\dagger_k \hc_{k+\omega/v}$ is the operator decreasing the electron wave vector of $\omega_0/v$ written in terms of the anticommuting fermionic operators $\hc_k$ and $\hc_k^\dagger$. The ladder operators $\ha_\omega$ and $\ha^\dagger_\omega$ respect the commuting relation $[\ha_\omega,\ha^\dagger_{\omega '}]=\delta(\omega-\omega')$. The coupling constant $g_\omega=\sqrt{\Gamma_{\rm EELS}(\omega)}$ dictating the rate of photons exchanged between electrons and the optical mode can be computed from the electron energy loss probability $\Gamma_{\rm EELS}(\omega)=(4e^2/\hbar)\int_{-\infty}^\infty dz dz' \cos[\omega(z-z')/v] {\rm Im}\{-G_{zz}(\Rb,z,\Rb,z',\omega)\}$ with the knowledge of the electromagnetic Green tensor $G(\rb,\rb',\omega)$ \cite{paper149}. We remind that the scattering operator in Eq. (\ref{scattSI}) links the joint electron-mode state in the interaction picture in the infinite past $\rho(-\infty)$ to the one after the interaction is ended $\rho(\infty)$ through the relation $\rho(\infty)=\hat{\mathcal{S}}(\infty,-\infty)\rho(-\infty)\hat{\mathcal{S}}^\dagger (\infty,-\infty)$.
In this work, we neglect the action of $\hat{\chi}$ on the electron bunch as it produces losses away from the optical mode frequencies and therefore could be filtered by energy spectrometer and because its effect influences only electrons which are temporally separated by few-fs whereas they are typically separated by hundreds of fs in bunches produced in transmission and scanning electron microscopes (SEM/TEM). Moreover, we consider the mode to have a high quality factor and to be well spectrally isolated from the other photonic resonances and having an electric field distribution $\vec{\mathcal{E}}_0(\rb)$. 

Under these condtions, we can approximate the Green tensor as $G_{zz}(\Rb,z,\Rb,z',\omega)=\mathcal{E}_{0,z}(\Rb,z)\mathcal{E}_{0,z}^*(\Rb,z')/2\pi \hbar \omega_0(\omega^2-\omega_0^2+\ii 0^+)$, with $0^+$ and infinitesimal positive number, which, plugged into the EELS probability allows us to rewrite the scattering operator as
\begin{align}
\hat{\mathcal{S}}(\infty,-\infty)\approx \ee^{\beta_0(\hb\ha^\dagger-\hb^\dagger \ha)},   \nonumber 
\end{align}
where, we have used the relation ${\rm Im}\{-1/(\omega^2-\omega_0^2+\ii 0^+)\}=\pi \delta(\omega-\omega_0)/2\omega_0$, we have defined the operators $\hb=\hb_{\omega_0}$, $\ha=\lim\limits_{\omega \to \omega_0}\ha_{\omega}/\sqrt{f(\omega-\omega_0)}$, with $f(\omega-\omega_0)={\rm Im}\{-1/\pi(\omega-\omega_0+\ii 0^+)\}$,  and we have introduced the single-mode coupling $\beta_0=(e/\hbar \omega_0)\big|\int_{-\infty}^\infty dz\,\mathcal{E}_{0,z}(\Rb,z)\ee^{-\ii\omega_0 z/v}\big|$. The commutation relation of the new bosonic operators can be computed through the limiting procedure $[\ha,\ha^\dagger]=\lim\limits_{\omega,\omega' \to \omega_0}[\ha_\omega ,\ha^\dagger_{\omega'}]/\sqrt{f(\omega-\omega_0)f(\omega'-\omega_0)}=1$.

We now write the $N$-electron the density matrix before entering in the interaction zone as $\rho_{e}(-\infty)=\sum_{\kb_N \kb_N'}\rho_{e,\kb_N,\kb_N'}|\kb_N\rangle \langle \kb_N'|$ by expanding its components in terms of $N$-dimensional vector states $|\kb_N\rangle=|k_1,\dots,k_N\rangle$ containing the longitudinal set of momenta of all electrons in the pulse. Then, we obtain the non-normalized post-interaction density matrix of the photonic mode, conditioned to the measurement $\qb_N$ of the final momenta of all electrons, by projecting the evolved joint density matrix onto the state $|\qb_N\rangle$, which reads
\begin{align}
T^{\qb_N}=\langle \qb_N| \ee^{\beta_0(\hb\ha^\dagger -\hb^\dagger \ha)} & \rho_{e}(-\infty)  \label{rhop1}\\
&\otimes  |\alpha\rangle \langle \alpha |\ee^{\beta_0(\hb^\dagger \ha-\hb\ha^\dagger)}|\qb_N \rangle,\nonumber
\end{align}
where we have also assumed the photonic mode to be previously coherently excited to the state $|\alpha\rangle=\ee^{-|\alpha|^2/2}\sum_n (\alpha^n/\sqrt{n!})|n\rangle$, i.e., we have taken $\rho(-\infty)=\rho_e(-\infty)\otimes |\alpha \rangle \langle \alpha|$. Eq. (\ref{rhop1}) can be reduced in a very simple form by noticing that the real-space state $|\zb_N\rangle=\sum_{\kb_N}(\ee^{-\ii \kb_N\cdot \zb_N}/L^{N/2})|\kb_N\rangle $ (with $L$ the quantization length) is an eigenstate of the electron destruction operator, i.e., $b|\zb_N\rangle=j(\zb_N)|\zb_N\rangle=\left(\sum_{i=1}^N\ee^{-\ii\omega_0z_i/v}\right)|\zb_N\rangle$, by using the normalization condition $\langle \zb_N |\zb'_N\rangle=\delta(\zb_N-\zb_N')$, and their completeness relation $\int d \zb_N |\zb_N\rangle \langle \zb_N|=\sum_{\kb_N} |\kb_N\rangle\langle \kb_N|=\mathcal{I}$. After some straightforward algebra involving the use of the property of the displacement operator $\ee^{\theta a^\dagger -\theta^* a}|\alpha\rangle = |\alpha+\theta\rangle$ from Eq. (\ref{rhop1}), we arrive at 
\begin{align}
T^{\qb_N}=\frac{1}{L^N}\!\!\int d\zb_N & d\zb'_N \,\rho_e (\zb_N,\zb_N')\,\ee^{\ii\qb_N\cdot (\zb_N'-\zb_N)} \label{rhop2} \\
&\times |\alpha +\beta_0 j(\zb_N)\rangle \langle \alpha +\beta_0 j(\zb_N')|,\nonumber 
\end{align}
where we have introduced the representation of the $N$-electron density matrix in space coordinates $\rho_e(\zb_N,\zb_N')=\langle \zb_N | \rho_e(-\infty)|\zb_N'\rangle=\sum_{\kb_N \kb_N'}\rho_{e,\kb_N \kb_N'}\ee^{\ii (\kb_N \cdot\zb_N-\kb_N'\cdot \zb_N')}/L^N$.

If we take our post-filtering procedure to be described by a function $F(\qb_N)$ integrating over only a finite set of prescribed electron momenta, we can retrieve final photonic state through the prescription $\rho_p=\sum_{\qb_N} F(\qb_N)T^{\qb_N}(\infty)/P_F=(L/2\pi)^N\int d\qb_N F(\qb_N)T^{\qb_N}(\infty)/P_F$, now normalized by the probability of successful filtering probability $P_F\leq 1$, which is given by Eq. (\ref{ls1}). Such form of the output light state can result quite useful when one is interested in the computation of photonic observables which can be written in terms of the normal ordered operators $a^{\dagger m} a^n$. For instance without post-filtering [$F(\qb_N)=1$], we can employ Eq. (\ref{ls1}) to compute $\langle a^{\dagger m} a^n\rangle ={\rm Tr}\{a^n \rho_p a^{\dagger m} \}=\int d\zb_N \rho_e(\zb_N,\zb_N)\beta_0^n(\zb_N) \beta_0^{* m}(\zb_N)$, which for $n=m=1$ reduces to the average number of emitted photons and agrees with the result in Ref. \citenum{paper373}. 

We can analyze two limits of Eq. (\ref{ls1}) depending on the shape of the filtering function: $(i)$ no post-sample filtering [$F(\qb_N)=1$], where $\rho_p$ only depends on the density $\rho_e(\zb_N,\zb_N)$ as already predicted in several other works \cite{RKT19,paper371,paper373}; $(ii)$ for a separable $N$-electron state $\rho_e(\zb_N,\zb_N')=\psi_e(\zb_N)\psi_e^*(\zb_N)$ and a filtering function well-peaked around a central value $\tilde{\qb}_{N}$, the photonic density matrix becomes a separable state, i.e., it factorizes as $\rho_p=|\psi_p\rangle \langle \psi_p|$ with 
\begin{align}
|\psi_p\rangle=\frac{f}{(2\pi)^{N/2}P^{1/2}_F}\int d\zb_N \psi_e(\zb_N)&\ee^{-\ii \tilde{\qb}_{N}\cdot \zb_N}\label{psipF}\\
&\times |\alpha +\beta_0(\zb_N)\rangle,\nonumber 
\end{align}
where $f=[\int d\qb_N F(\qb_N)]^{1/2}$. Equation (\ref{psipF}) states that a perfect energy post-filtering performed on a pure $N$-electron state yields a pure photonic state.

\section{Number-state representation of the photonic density matrix: multi-electron Wigner function and the projected coherence factor (PCF)}
\label{secB}
\renewcommand{\theequation}{B\arabic{equation}}

\begin{figure*}
\includegraphics[scale=1]{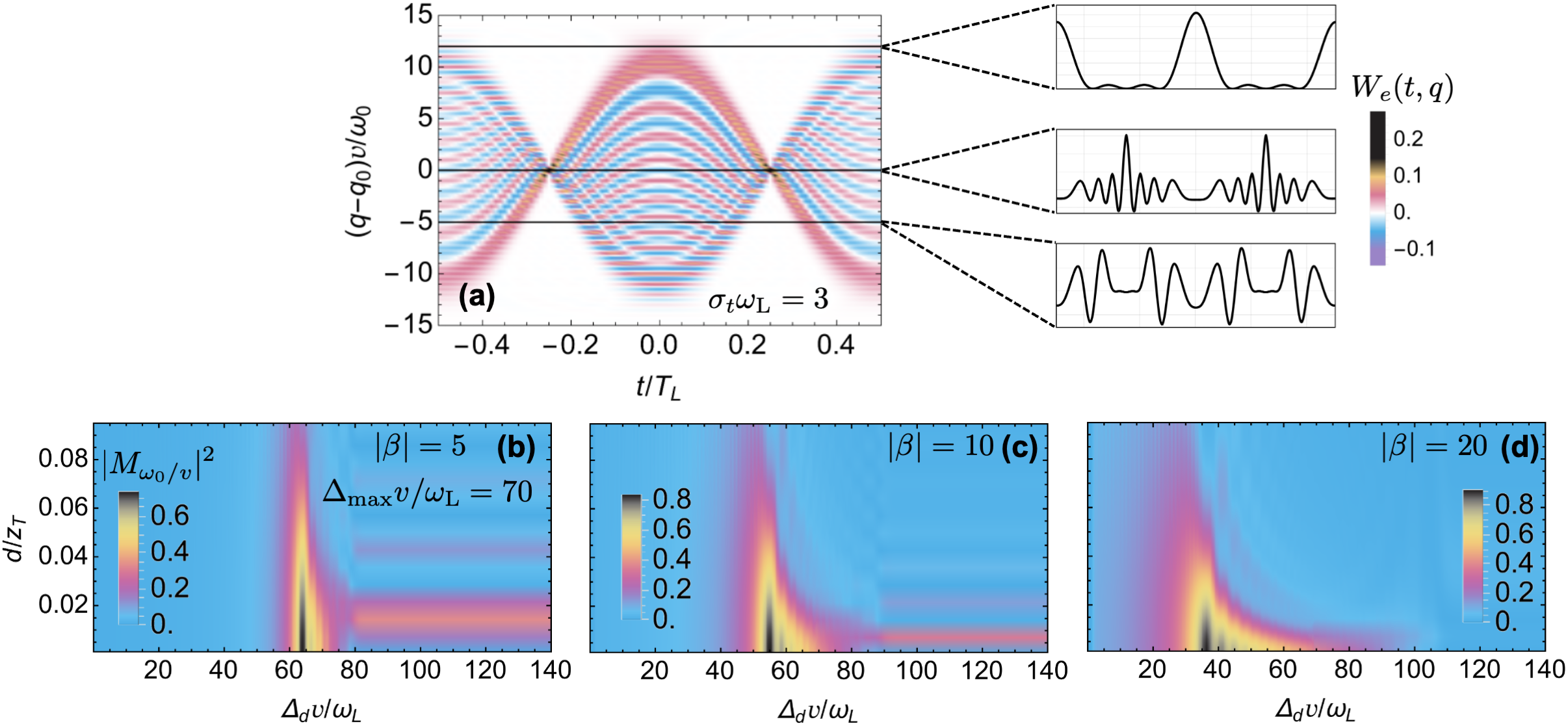}
\caption{{\bf Coherence of modulated electrons}. {\bf (a)} Momentum-time correlation expressed by the electron Wigner function [see Eq. (\ref{wie}) with $t=z/v$ and $T_{\rm L}=2\pi/\omega_{\rm L}$] with coherence time $\sigma_t\omega_{\rm L}=3$ right after a IELS modulation $c_\ell=J_\ell(2|\beta|)\ee^{\ii \ell {\rm arg}\{-\beta\}}$ of interaction strength $|\beta|=5.7$ at laser frequency $\omega_{\rm L}=\omega_0$ . The cuts along the time axis show a well-defined sub-cycle modulation for fixed normalized momentum. {\bf (b-d)} Absolute squared value of the CF $|M_{\omega_0/v}|^2$ for an electron after an IELS interaction, free propagation of a distance $d$ [appending a phase $-2\pi \ell^2 d/z_{\rm T}$ to the $c_\ell$ used in (a)], and an energy filtering stage [see Eq. (\ref{Mfilt})] selecting only a portion of longitudinal momenta $\Delta_d$ for $|\beta|=5$ (b), $10$ (c), and $20$ (d). The maximum values found are $|M_{\omega_0/v}|^2\sim 0.74 $, 0.84, and 0.91 respectively. In all panels ${\rm arg}\{-\beta\}=0$.}
\label{FigS1}
\end{figure*}

We want now to isolate the contribution of the electron state to the formation of $\rho_p$. In order to do that, we study the number representation of the photonic density matrix $\rho_p=\sum_{n,n'=0}^\infty \rho_{p,nn'}|n\rangle \langle n' |$ for a generic $N$-electron state and post-filtering operation. With the aid of the multinomial equality $\big(\sum_{i=1}^N x_i\big)^n\big(\sum_{i=1}^N y_i\big)^k=\sum_{\mb,\mb' \geq 0} \,C^{(n,k)}_{\mb \mb'}\prod_{i=1}^N x_i^{m_i}y_i^{m_i'}$, written in terms of the coefficient $C^{(n,k)}_{\mb \mb'}=(n;m_1',\dots,m_N')(k;m_1'',\dots,m_N'')$ and the multinomial factors $(n;m_1,\dots,m_N)=(n!/m_1!\dots m_N!)$, with the superscript $(n,k)$ restricting (the coefficients are imposed to vanish otherwise) the sum over $\mb$ ($\mb'$) to the combinations satisfying $m_1+\dots+m_N=n$ ($m'_1+\dots+m'_N=k$), we rewrite the components of Eq. (\ref{ls1}) for $\alpha=0$ as
\begin{align}
&\rho_{p,nn'}=\frac{1}{P_F} \sum_{\substack{k,k',\mb\\ \mb',\pb,\pb' \geq 0}}\mathcal{C}_{\mb\mb'\pb\pb'}^{(n,k,n',k')} \int d\qb_N F(\qb_N)\label{rhonnpNSI} \\
& \!\! \!\! \times \PM_{\omega_0(\mb'-\mb+\pb-\pb')/v}\Big[\qb_N+\frac{\omega_0}{2v}(\mb-\mb'+\pb-\pb')\Big],\nonumber
\end{align}
where we have introduced the $\beta_0$-dependent combinatorial coefficient 
\begin{align}
\mathcal{C}_{\mb\mb'\pb\pb'}^{(n,k,n',k')}=[(-2)^{-(k+k')}&\beta_0^{2(k+k')+n+n'}/n!n'!k!k'!]\nonumber \\
&\quad  \times C^{(n+k,k)}_{\mb\mb'}C^{(n'+k',k')}_{\pb\pb'}. \label{combc} 
\end{align} 
Equation (\ref{rhonnpNSI}) shows that the $N$-electron density matrix appears only in terms of the projected coherence factor (PCF)
\begin{align}
\PM_{\kb_N}(\qb_N)= \int d\zb_N W_e(\zb_N,\qb_N)\,\ee^{\ii \kb_N\cdot \zb_N}\label{pcfSI}
\end{align}
defined through the quantum generalization of the classical phase-space density for the multi-electron state: the $N$-electron Wigner function 
\begin{align}
W_e&(\zb_N,\qb_N)=\label{wignerf}\\
&\frac{1}{(2\pi)^N} \int d\yb_N \rho_e(\zb_N-\yb_N/2,\zb_N+\yb_N/2)\,\ee^{\ii \qb_N\cdot \yb_N}.\nonumber
\end{align}
The term PCF is inspired by the coherence factor (CF) defined in several other works \cite{PG18,paper373,paper374} for a single 
particle in an electron bunch of uncorrelated electrons as the Fourier transform of the density $M_{k}=\int_{-\infty}^\infty dz\,\rho_e(z,z)\,\ee^{\ii k z}$, to which it reduces when no post-filtering is performed. This last statement can be simply verified by integrating Eq. (\ref{pcfSI}) over $\qb_N$ and by using the property of the Wigner function $\int d\qb_N W_e(\zb_N,\qb_N)=\rho_e(\zb_N,\zb_N)$. The PCF contains the Fourier components of the Wigner function for a given post-selected longitudinal momentum window. In Fig. (\ref{FigS1}a), we report some cuts of the Wigner function for a single electron integrated over an infinitesimal momentum window [see Eq. (\ref{wie}) below]. 

In the case of uncorrelated electrons, the density matrix can be written as the product of one-electron density matrices $\rho_e(\zb_N,\zb_N')=\prod_{i=1}^N \rho_e^i(z_i,z'_i)$, which in turn, given Eq. (\ref{pcfSI}) and Eq. (\ref{wignerf}), leads to the factorization of the PCF $\PM_{\kb_N}(\qb_N)=\prod_{i=1}^N \PM^i_{k_i}(q_i)$.
Moreover, in the special case of pure electron states $\rho_e^i(z,z')=\psi^{i}_e(z)\psi^{i*}_e(z')$ and of a post-filtering window narrow around the vector $\tilde{\qb}_{N}=\omega_0 \sb/v$, with $\sb$ and $N$-dimensional vector of integer numbers, the state in Eq. (\ref{rhonnpNSI}) purifies and the state coefficients of Eq. (\ref{psipF}) become
\begin{align}
\alpha_{p,n}=\frac{f}{P_F^{1/2}}&\sum_{k,\mb,\mb'\geq0}[C_{\mb\mb'}^{(n+k,k)}(-2)^{-k}\beta_0^{2k+n}/n!k!] \nonumber \\
&\times \prod_{i=1}^N\int_{-\infty}^\infty \!\! \frac{dz}{\sqrt{2\pi}}\, \psi^i_e(z)\ee^{\ii \omega_0(m_i'-m_i+s_i)z/v}.\nonumber
\end{align}
The previous expression assumes a useful form for the application of the modulation optimization algorithm presented in Sec. \ref{secD} to multiple electrons having a wave packet with infinite coherence time $\sigma_t=L/v$ (where $L\rightarrow \infty$ at the end of the calculations) of the type $\psi_e^i(z)=\sum_{\ell=-\infty}^\infty c_\ell^i\,\ee^{\ii \ell \omega_0  z/v}/\sqrt{L}$. Indeed, by taking $F(\qb)=(2\pi/L)^N\delta(\qb-\tilde{\qb}_{N})$ and such type of the state, the coefficients $\alpha_{p,n}$ in the number representation $|\psi_p\rangle =\sum_{n=0}^\infty \alpha_{p,n}|n\rangle $ becomes 
\begin{align}
    \alpha_{p,n}=\frac{1}{P_F^{1/2}}\!\!\!\sum_{k,\mb,\mb'\geq 0}[(-2)^{-k}&\beta_0^{2k+n}C^{(n+k,k)}_{\mb \mb'}/n!k!]\nonumber \\
    &\times \prod_{i=1}^N c^i_{m_i+s_i-m_i'}.\label{sealpha}
\end{align}

\subsection{PCF for electrons with stochastic arrival times}
In SEM/TEM, the coherence time of each electron $\sigma_t$ is typically several times smaller than its classical (or incoherent) uncertainty $\Delta_t$ acquired by the electron ensemble through the random fluctuations of the electron source and of the instrumentation. Such fluctuations are responsible for the random arrival times at which the electrons reach the sample plane. In order to explore the consequences linked to this incoherent portion of the $N$-electron state, we study uncorrelated electrons with density matrix $\rho_e^i(z,z')=\int_{-\infty}^\infty dz_0 P(z_0) \psi_e^i(z,z_0)\psi_e^{i*}(z',z_0)$ defined by a classical distribution $P(z_0)$ of longitudinal planes $z_0$ that the electron crosses at $t=0$ and a coherent wave function $\psi^i_e(z,z_0)$. For instance, in the case of an electron modulated by a IELS interaction at frequency $\omega_{\rm L}$, for which the wave packet takes the general form $\psi_e^i(z,z_0)=\psi^i_0(z,z_0)\sum_{\ell=-\infty}^\infty c_\ell^i \,\ee^{\ii  \ell \omega_{\rm L} z/v}$  assuming a Gaussian envelope $\psi^i_0(z,z_0)=\ee^{-(z-z_0)^2/4v^2\sigma_t^2+\ii q_0 z}/(2\pi v^2\sigma_t^2)^{1/4}$, with the coefficients $c^i_\ell$ which depends on the form of modulation [see for instance Eq. (\ref{pinem1})] and they are chosen to ensure the normalization condition $\int_{-\infty}^\infty dz|\psi^i_e(z,z_0)|^2=1$.

By plugging this density matrix in the definition of the Wigner function and the PCF, one obtains
\begin{subequations}
\label{wiepmie}
\begin{align}
&W^i_e(z,q)=\frac{1}{\pi}\sum_{\ell,\ell'}c_\ell c_{\ell'}^*\, \label{wie}\\
&\quad \times \ee^{-[q-q_0-(\ell+\ell')\omega_{\rm L}/2v]^2 2v^2\sigma_t^2}\ee^{\ii (\ell-\ell')\omega_{\rm L} z/v} \nonumber \\
&\quad \times \int_{-\infty}^\infty \!\!\!dz_0 P(z_0)\ee^{-(z-z_0)^2/2v^2\sigma_t^2},\nonumber \\
&\PM_k^i(q)=\sqrt{\frac{2v^2\sigma_t^2}{\pi}}\sum_{\ell,\ell'=-\infty}^\infty c_\ell c_{\ell'}^*\label{pmie}\\
&\quad\times\ee^{-[q-q_0-(\ell+\ell')\omega_{\rm L}/2v]^2 2v^2\sigma_t^2} \ee^{-[(\ell-\ell')\omega_{\rm L}/v+k]^2v^2\sigma_t^2/2}\nonumber \\
&\quad \times \int_{-\infty}^\infty \!\!\!dz_0 P(z_0)\ee^{\ii [(\ell-\ell')\omega_{\rm L}/v+k] z_0} \nonumber .
\end{align}
\end{subequations}

For momenta $k=m\omega_0/v$ and modulation frequency $\omega_{\rm L}=\omega_0$, with $m$ and integer, as required by the computation of Eq. (\ref{rhonnpNSI}), and in the limit of $\sigma_t \omega_0\gg 1$, the effect of the exponential in the first line of Eq. (\ref{pmie}) and the one arising from the incoherent integral $\int_{-\infty}^\infty \!\!\!dz_0 P(z_0)\ee^{\ii [(\ell-\ell')\omega_0/v+k] z_0}$ is equivalent, i.e., to enforce the $m=\ell'-\ell$ condition. Indeed, for a Gaussian ensemble of arrival times $P(z_0)=\ee^{-z_0^2/2v^2\Delta t^2 }/\sqrt{2\pi v^2 \Delta t^2}$ where typically $\Delta t \omega_0\gg \sigma_t \omega_0 \gg 1$, such integral gives $\ee^{-[(\ell-\ell')\omega_0/v+k]^2v^2\Delta t^2/2}$. Therefore in this regime, $\rho_p$ can be equivalently evaluated by directly starting from the pure single-electron density matrix $\rho^i_e(z,z')=\psi^i_e(z)\psi^{i *}_e(z')$ disregarding the incoherent average on $z_0$. However, the purity of light states generated by electrons with coherence times smaller than the mode optical cycle will be strongly affected by it.
\subsection{Intensity fluctuations generated by $N$ uncorrelated electrons}
From Eq. (\ref{ls1}), we can compute the amount of light emitted $I_N=\langle \ha^\dagger \ha \rangle\equiv  \langle \hn \rangle$ and its fluctuations $\Delta I_N=\langle \hn^2 \rangle- \langle \hn \rangle^2$ by $N$ modulated electrons with random times of arrival and large coherence times. This is easily done by utilizing the properties of the coherence state to obtain
\begin{align}
I_N=\beta_0^2\bigg[N+\sum_{i\neq i'}^N M^i_{\omega_0/v} M^{i' *}_{\omega_0/v}\bigg],\nonumber \\
\Delta I_N^2/I_N = 1 + I_N[\mathcal{G}_N/I_N^2-1],\nonumber 
\end{align}
where we have defined $\mathcal{G}_N=\langle \ha^{\dagger 2}\ha^{ 2}\rangle=\beta_0^4 \sum_{\mb,\mb'\geq 0} C^{(2,2)}_{\mb,\mb'} \prod_{i=1}^N M^i_{\omega_0(m_i'-m_i)/v}$.
Super-Poissonian statistics is observed for electrons modulated such that $I_N^2<\mathcal{G}_N$. Reassuringly, by exploiting the property $\sum_{\mb\geq 0}\delta_{m_1+\dots +m_N,n}(n;m_1,\dots,m_N)=N^n$, we recover a Poissonian emission $I_N=\Delta I^2_N$ in the limit of classical electrons for which $M_{\omega_0 m/v}=1$ for any $m$. Interestingly, since $\Delta I_N^2$ and $I_N$ must be real positive numbers and the ratio $\mathcal{G}_N/I_N^2$ does not depend on $\beta_0$, we conclude that CF yielding $\mathcal{G}_N/I_N^2<1$ would lead to arbitrary negative fluctuations for an increasing spontaneous emission coupling thus corresponding to unphysical electron states. 

\section{Mode density matrix after the interaction with a single electron}
\label{secC}
\renewcommand{\theequation}{C\arabic{equation}}

When only a single modulated electron is involved, we have $C^{(n+k,k)}_{mm'}=C^{(n'+k',k')}_{pp'}=1$ which directly allows us to rewrite Eq. (\ref{rhonnpNSI}) as
\begin{align}
&\rho_{p,nn'}=\frac{1}{P_F} \langle n|\beta_0\rangle \langle \beta_0|n'\rangle\label{state1elSI}\\
&\quad\times \int_{-\infty}^\infty  dq\,F(q) \,\PM_{\omega_0(n'-n)/v}[q+\omega_0(n+n')/2v].\nonumber 
\end{align}
For a post-filtering close to the $m$-th sideband, we can take $F(q)$ to vanish everywhere apart from the segment $q_0+s\omega_0/v+[-\delta_d,\delta_d]$, that, plugged into Eq. (\ref{state1elSI}) with the electron state used to obtain Eq. (\ref{pmie}) modulated at frequency $\omega_{\rm L}=\omega_0$, and with a Gaussian incoherent ensemble, yields
\begin{align}
\rho_{p,nn'}=&\frac{1}{2 P_F} \langle n|\beta_0\rangle \langle \beta_0|n'\rangle \label{state2el}\\
&\times \sum_{\ell,\ell'=-\infty}^\infty c_\ell c_{\ell'}^* \,\ee^{-[\ell-\ell'+n'-n]^2\omega_0^2(\sigma_t^2+\Delta t^2)/2}\nonumber \\
&\times \Big\{{\rm Erf}\big[\sqrt{2}\omega_0\sigma_t (\delta_d v/\omega_0+x_0)\big]\nonumber 
\\ &\quad \quad \quad \quad \quad \quad +{\rm Erf}\big[\sqrt{2}\omega_0 \sigma_t(\delta_d v/\omega_0-x_0) \big]\Big\},\nonumber
\end{align}
where we have made used of the integral $\int_{-\delta_d}^{\delta_d} dx \exp{-(x-x_0)^2\sigma^2}=\sqrt{\pi/4\sigma^2}\{{\rm Erf}[(\delta_d-x_0)\sigma]+{\rm Erf}[(\delta_d+x_0)\sigma]\}$ with $x_0=(\ell+\ell')/2 - [(n + n')/2 +s]$ and $\sigma=\sqrt{2}\omega_0\sigma_t $. The function ${\rm Erf}(x)=(2/\sqrt{\pi})\int_0^x dz \, \ee^{-z^2}$ is the error function. In the limit $\delta_d \sigma_t v \gg 1$, one can verify that the state only depends on the CF $M_{\omega_0(n'-n)/v}=\sum_{\ell,\ell'} c_\ell c_{\ell'}^* \exp{-(\ell-\ell'+n'-n)^2\omega_0^2(\sigma_t^2+\Delta t^2)/2}$  and $M_{\omega_0(n'-n)/v}\approx \sum_{\ell=-\infty}^\infty c_\ell c_{\ell+n'-n}^*$ for $\sqrt{\sigma_t^2+\Delta t^2} \omega_0 \gg 1$. In the opposite limit of precise sideband determination ($\delta_d \sigma_t v \ll 1$), by using the expansion ${\rm Erf}[\sigma(x+x_0)]+{\rm Erf}[\sigma(x-x_0)]\approx (4 \sigma x/\sqrt{\pi}) \exp{-x_0^2\sigma^2}$, we obtain the separable state $\rho_p=|\psi_p\rangle\langle \psi_p|$ if $\Delta t \omega_0 \ll 1$ or $\sigma_t \omega_0 \gg 1$. 

From Eq. (\ref{state2el}), in the case of large coherence time ($\sigma_t \omega_0 \gg 1$) and perfect post-filtering procedure ($\delta_d \sigma_t v \ll 1$), we obtain a pure state [in agreement with Eq. (\ref{sealpha})] with expansion coefficients
\begin{align}
\alpha_{p,n}=\frac{1}{P_F^{1/2}}\langle n|\beta_0\rangle\, c_{n+s},\label{alphapend1}
\end{align}
where $P_F=\sum_{n=0}^\infty |\langle n|\beta_0\rangle|^2 |c_{n+s}|^2$. It is interesting to notice that, since the normalization constant $P_F\leq 1$ and the average number of photons respects the inequality $\sum_{n=0}^\infty n |\alpha_{p,n}|^2\leq \beta_0^2/P_F$, its value can assume values larger than the number of photons one would measure without post-filtering the electron energy. Meaningfully, because $P_F$ represents the probability of such post-filtering procedure, the larger the deviation from the average, the bigger the time needed to acquire sufficient statistics. An evident constraint arising from Eq. (\ref{alphapend1}), it is related to the asymptotic behavior of $\alpha_{p,n}$. Indeed, since the electron coefficients are normalized ($\sum_{\ell=-\infty}^\infty |c_\ell|^2=1$), the limit $\lim_{n\rightarrow\infty}\alpha_{p,n}/\langle n|\beta_0\rangle=0$ needs to be satisfied for the electron state to be physical. This restricts the possible syntheses to states which have any type of coefficient over a finite set of $\alpha_{p,n}$, for instance by choosing all values from $\alpha_{p,0}$ to $\alpha_{p,n_{\rm max}}$, and then which decay faster than the components of a coherent state. Due to its generality, this procedure allows for almost perfect generation of any type of state as long as its average number of photons is $\ll n_{\rm max}$.  

\subsection{Coherence factor of a modulated electron after energy filtering}
We want now to analyze the CF $M_k=\int_{-\infty}^\infty dz \rho_e(z,z) \,\ee^{\ii k z}/M_0$ (the factor $M_0$ has been added for normalizing the electron density matrix), for a modulated Gaussian electron at the exit of an energy filter \cite{TI13}. In order to do it, we firstly need to compute the electron state after the filtering process which we write by taking the Fourier components of $\rho_e(z,z')=\int_{-\infty}^\infty dq dq' \,\rho_e(q,q')\,\ee^{\ii q z-\ii q' z'}/4\pi^2$ and then by multiplying them by a function $\mathcal{W}(q)$ representing the energy-filtering process. 

It is convenient to evaluate the CF through the expression $M_k=\int_{-\infty}^\infty dq\, \rho_e(q,q+k) \mathcal{W}(q) \mathcal{W}(k+q)/2\pi M_0$, which, for $\mathcal{W}(q)=\theta(q-q_0-\Delta_{\rm max}+\Delta_d)\theta(\Delta_{\rm max}-q+q_0)$, with $\Delta_d>0$, selecting longitudinal momenta in the range $[\Delta_{\rm max}-\Delta_d,\Delta_{\rm max}]$ around $q_0$ for the electron state used to write Eq. (\ref{state2el}), gives
\begin{align}
M_k&=\frac{1}{2M_0}\theta[\Delta_d-|k|]\sum_{\ell,\ell'}c_\ell c_{\ell'}^*\label{Mfilt} \\
&\times \ee^{-(\ell-\ell'+vk/\omega_{0})^2(\sigma_t^2+\Delta t^2) \omega_0^2/2} \nonumber \\
&\times  \Big\{ {\rm Erf}\big[(2\Delta_{\rm max}-2k_+ -  k_{\ell+\ell'})\sigma_t\omega_0/\sqrt{2}\big]\nonumber \\
&~~ +{\rm Erf}\big[( k_{\ell+\ell'}+2k_--2\Delta_{\rm max}+2\Delta_d)\sigma_t\omega_0/\sqrt{2}\big]\Big\},\nonumber 
\end{align}
where $k_{\ell+\ell'}=(\ell+\ell')\omega_{0}/v-k$, $k_+={\rm max}\{0,k\}$, and $k_-={\rm min}\{0,k\}$. In the $\sigma_t \omega_0\gg 1$ limit, the CF of Eq. (\ref{Mfilt}) at $k=m\omega_0/v$, for $m_{\rm max}=\lfloor \Delta_{\rm max}v/\omega_0\rfloor $ and $m_{	\rm min}=\lfloor (\Delta_{\rm max}-\Delta_d)v/\omega_0 \rfloor$, where $\lfloor x\rfloor $ returns the integer part of $x$, reduces to 
\begin{align}
M_{\omega_0m/v}=\frac{1}{M_0}\sum^{m_{\rm max}-{\rm max}\{0,m\}}_{\ell=m_{\rm min}-{\rm min}\{0,m\}}c_\ell c^*_{\ell +m}.\label{filtM}
\end{align} 
Interestingly, this filtering procedure can lead to CF of larger absolute values than the unfiltered version but the number of included energy coefficients needs to be larger than the order $m$ for the CF to do not vanish, i.e., $m_{\rm max}-m_{\rm min}\geq |m|$ [see Fig. (\ref{FigS1}b-c)]. Pre-sample filtering is intimately connected to post-filtering as the CF of a filtered electron can be rewritten in terms of the PCF of an unfiltered electron $\PM^{{\rm unf}}_k(q+k/2)=\rho_e(q,q+k)/2\pi$ as $M_k=\int_{-\infty}^\infty dq \mathcal{W}(q)\mathcal{W}(q+k) \PM^{{\rm unf}}_k(q+k/2)/M_0$.

\subsection{Natural synthesis of cat states after a single unstructured inelastic electron-light scattering (IELS) interaction}

\begin{figure*}
\includegraphics[scale=1.1]{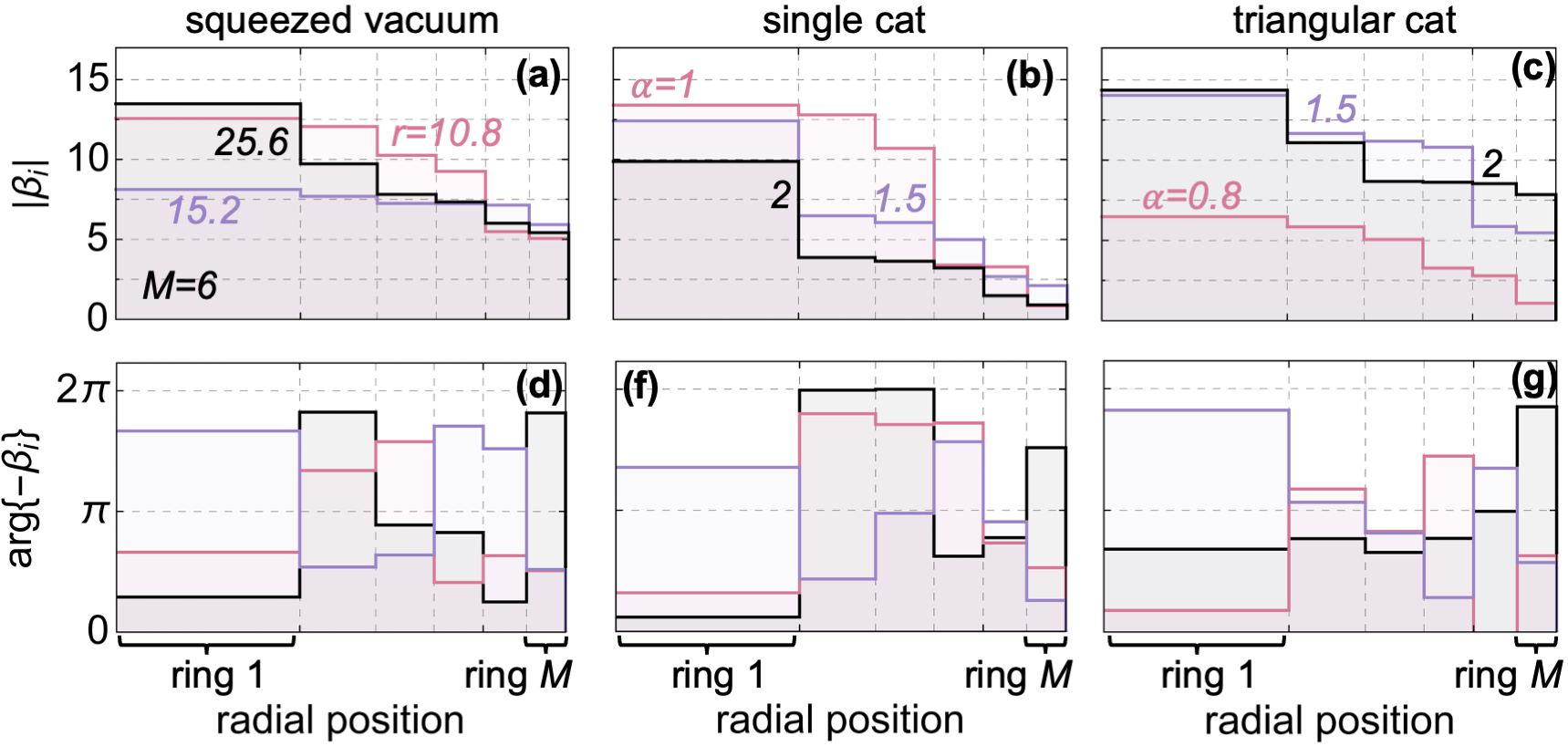}
\caption{{\bf IELS coupling strength profiles}. Optimal coupling coefficients $\beta_i$ for the cases signaled by the black dots in Fig. (\ref{Fig4}b-d). Panels {\bf (a-c)} reports their amplitudes whereas panels {\bf (a-g)} their phase.}
\label{FigS2}
\end{figure*}

When an electron traverses an electric field distribution $\vec{\mathcal{E}}(\rb,t)=\vec{\mathcal{E}}(\rb)\ee^{-\ii \omega_{\rm L} t}+{\rm c.c.}$, arising from the scattering of a laser pulse of photon energy $\hbar \omega_{\rm L}$ onto a nanostructure, its initial wave function $\psi_{0}(z,t)$ undergoes inelastic electron-light scattering (IELS) modifying its spatial and energetic structure. For relativistic electrons, the exact exit state after traveling a distance $d$ comparable with the Talbot distance $z_{\rm T}=4\pi\me v^3\gamma^3/\hbar \omega_{\rm L}$, within the electron conditions considered in this work, can be found in several works \cite{PLZ10,paper360,paper371} and reads
\begin{align}
\psi_{\rm IELS}(z,t)=&\psi_0(z,t)\sum_{\ell=-\infty}^\infty J_\ell(2|\beta|)\label{pinem1}\\
&\times \ee^{\ii \ell \omega_0(z-vt)/v+\ii \ell {\rm arg}\{-\beta\}-2\pi \ii \ell^2 d/z_{\rm T}},\nonumber
\end{align}
where $J_\ell(x)$ is the $\ell$-th other Bessel function, $\beta=(e/\hbar \omega_{\rm L})\int_{-\infty}^\infty dz\, \mathcal{E}_{z}(\Rb,z)\,\ee^{-\ii \omega_{\rm L} z/v}$. By comparing the electron wave function used to obtain Eq. (\ref{wie}) and   Eq. (\ref{pinem1}), we identify the coefficients $c_n$ typical of a IELS interaction, which therefore read  $c_{n}=J_{n}(2|\beta|)\,\ee^{\ii n{\rm arg}\{-\beta\}}$. 
Now, we want to study the effect of the $n_{\rm max}$ IELS electron coefficients $c_s,\dots,c_{s+n_{\rm max}}$ on the generated light state. In order to do that, we assume them to vanish for $n>s+n_{\rm max}$, namely we take
\begin{align}
\!\!\!c_{n}\!\!=\!\!\begin{cases} \frac{1}{\mathcal{N}^2}J_{n}(2|\beta|)\ee^{\ii n{\rm arg}\{-\beta\}},\,s\!\leq n \! \leq \!s+n_{\rm max},\\
0,\quad ~~\, \quad \quad\quad\quad\quad \quad ~~\,\,{\rm otherwise},
\end{cases}\label{pinemfilter}
\end{align}
where $\mathcal{N}^2=\sum_{m=s}^{s+n_{\rm max}} J^2_{m}(2|\beta|)$ is a normalization constant. 

For a very high electron-light coupling, we can take the asymptotic expansion  $J_n(2|\beta|)\approx (\pi|\beta|)^{-1/2}\cos[2|\beta|-n\pi/2-\pi/4]$ \cite{AS1972} valid for filtering values $(n_{\rm max}+s)^2\ll 2|\beta|$. By plugging the previous expression into Eq. (\ref{pinemfilter}) with $k=0$ and working out the normalization factor $P_F$ with the help of the relation $\sum_{k=0}^n\lambda^k/k!=\ee^\lambda\, \Gamma(n+1,\lambda)/n!$, where $\Gamma(n,x)=\int_x^\infty t^{n-1}\ee^{-t}dt$ is the incomplete gamma function, we obtain
\begin{align}
\!\!\alpha_{p,n}\!=\!\begin{cases}\frac{\langle n| \chi\rangle}{P^{1/2}_F}\big[1\!+\!\ee^{\ii(s \pi+\pi/2-4|\beta|)}(-1)^n\big],\,0\leq n \leq n_{\rm max},\\
0,\quad \quad \quad \quad \quad\quad\quad\quad\quad\quad\quad\quad ~~{\rm otherwise},
\end{cases}
\label{cat1el}
\end{align}
where $\chi=-\ii \beta_0\ee^{\ii {\rm arg}\{-\beta\}}$ the dividing factor can now be written in the compact form $P_F=2\big[\Gamma(n_{\rm max}+1,\beta_0^2)+(-1)^s\ee^{-2\beta_0^2}\sin(4|\beta|)\Gamma(n_{\rm max}+1,-\beta_0^2)\big]/n_{\rm max}!$. Eq. (\ref{cat1el}) needs to be compared with the photon-number coefficients of a cat state $\langle n|{\rm cat}_\theta^\alpha\rangle=\langle n|\alpha \rangle[1+\ee^{\ii \theta}(-1)^n]/[2+2\cos(\theta)\ee^{-2|\alpha|^2}]^{1/2}$ to realize that a cat state with $\theta=s\pi+\pi/2-4|\beta|$ and $\alpha=\chi$ is created by a single electron-light modulation, filtering, and post-filtering with a precision depending on the value of $s+n_{\rm max}$. 

\section{Modulation of energy coefficients through laterally-structured IELS interaction and their optimization}
\label{secD}
\subsection{Energy coefficients in the interaction region}

\renewcommand{\theequation}{D\arabic{equation}}

In this section, we report a variation of the method presented in Ref. \cite{paper415} to produce approximated electron energy coefficients $c_\ell$ as close as possible to the ones needed to crate a given target light state $\alpha^{\rm targ}_n$, according to the relation in Eq. (\ref{alphapend1}). This method leverages a wide electron beam traversing a near-field structured in concentric circular sections [see sketch in Fig. (\ref{Fig4}a)] at plane $z=0$ which is then focused to the focal point $(\Rb,z)=(0,z_0+f)$ by an axially symmetric and aberration-free converging lens placed at $z_0$, with radius $R_{\rm max}$ and numerical aperture $\NA\approx R_{\rm max}/f$. 

The time-dependent electron wave after passing through such interaction can be written as the three-dimensional extension of Eq. (\ref{pinem1}) \cite{paper415}
\begin{align}
\psi_{\rm IELS}(\rb,t)&= \psi_0(\rb,t)\, \ee^{-\ii q_0 z}\label{IELS2}\\
&\times \sum_{\ell=-\infty}^\infty J_\ell(2|\beta(\Rb)|)\ee^{\ii q_\ell z+\ii {\rm arg}\{-\beta(\Rb)\} -\ii \ell \omega_{\rm L} t} \nonumber,
\end{align}
where we have introduced the longitudinal momentum $q_\ell\approx q_0+\ell \omega_{\rm L}/v-\ell^2/z_{\rm T}$ corresponding to an energy $E^e_0 +\hbar \omega_{\rm L}\ell$. If we assume the electron to be well collimated and covering the entire extension of the interaction zone, we can take $\psi_0(\rb,t)=\psi_0\, \ee^{\ii q_0 z -\ii E^e_0  t/\hbar }$. The action of the converging lens can be expressed by multiplying energy amplitude of Eq. (\ref{IELS2}) with $\theta(R_{\rm max}\!\! -\!\!R)\ee^{-\ii q_\ell R^2/2 f}$ which at the lens' plane becomes  
\begin{align}
\psi^{\rm lens}_{\rm IELS}(\Rb,z_0,t)=& \psi_0 \,\ee^{-\ii E^e_0  t }\sum_{\ell=-\infty}^\infty  J_\ell[2|\beta(\Rb)|]\label{IELSl}\\
&\times \ee^{\ii q_\ell z_0+\ii {\rm arg}\{-\beta(\Rb)\} -\ii \ell \omega_{\rm L} t} \ee^{-\ii q_\ell R^2 /2f}.\nonumber
\end{align}
Now, we use scalar diffraction theory \cite{B1954,MW95} to propagate the wave function of Eq. (\ref{IELSl}) from the plane $z_0$ to the focal plane $z_0+f$. Indeed, from the knowledge of an electron wave $\psi_\ell(\Rb,z_0)$ with total momentum $q_\ell$ at $z_0$, $\psi_\ell(\Rb,z_s)$ can be obtained through the expression
\begin{align}
\psi_\ell(\Rb,z_s)&=\frac{1}{(2\pi)^2}\int d^2\Qb\,\ee^{\ii \Qb\cdot \Rb + \ii q_z^\ell (z_s-z_0)} \nonumber \\
&\quad \quad \quad \times \int d^2 \Rb' \psi_\ell(\Rb',z_0)\,\ee^{-\ii \Qb\cdot \Rb'}\nonumber \\
& \approx \frac{-\ii q_\ell }{2\pi(z_s-z_0)} \int d^2 \Rb' \psi_\ell(\Rb',z_0)\label{scal} \\
&\quad\quad\quad  \times \ee^{\ii |\Rb-\Rb'|^2 q_\ell/2(z_s-z_0)+\ii q_\ell (z_s-z_0)},\nonumber
\end{align} 
where the last line was obtained by taking the paraxial approximation $q_z^\ell=\sqrt{q_\ell -Q^2}\approx q_\ell -Q^2/2q_\ell$ and the integral $\int_0^\infty dx\, x\,\ee^{ -\ii a x^2}J_0(bx)=(-\ii/2a)\ee^{\ii b^2/4a}$ [Eq. 6.631-4 of Ref. \citenum{GR1980}]. By applying Eq. (\ref{scal}) to each energy component of Eq. (\ref{IELSl}) and by employing the axial symmetry of the field, that implies $\beta(\Rb)\equiv \beta(R)$, one arrives to the expression
\begin{subequations}
\begin{align}
\psi^{\rm lens}_{\rm IELS}(\Rb,z_s,t)&=\frac{- \ii \psi_0 f^2}{(z_s-z_0)}\ee^{- \ii E^e_0  t/\hbar} \label{ielslwf}\\
\times& \sum_{\ell=-\infty}^\infty b_\ell(\Rb,z_s-z_0)\, \ee^{\ii q_\ell z_s-\ii \ell \omega_{\rm L} t},\nonumber \\
b_\ell(\Rb,z_s-z_0)&=q_\ell \,\ee^{ \ii R^2 q_\ell/2(z_s-z_0)}\label{ielslcc} \\
\times& \int_0^{\NA} \!\!\! \!\!\! d\theta\,\theta\, J_0\Bigg(\frac{R f q_\ell \theta }{z_s-z_0}\Bigg) J_\ell[2|\beta(\theta)|]\nonumber \\
\times& \ee^{-\ii \theta^2 q_\ell f (z_s-z_0-f)/2(z_s-z_0)}\ee^{\ii \ell{\rm arg}\{-\beta(\theta)\}}. \nonumber 
\end{align}
\end{subequations}
Since we are interested in the electron wave function close to interaction with the cavity, assumed to be placed at the focus, and since the coefficients $c_\ell(\Rb,z_s-z_0)$ do not vary considerably along its extension $\sim 100$ $\mu$m for electron kinetic energies $\sim 100$ keV, $\NA\sim 2\times 10^{-4}$, we take $b_\ell(\Rb,z_s-z_0)\approx b_\ell(\Rb=0,f)$ in Eq. (\ref{ielslwf}). In addition, by approximating $q_\ell$ with its second order Taylor expansion in the exponential of Eq. (\ref{ielslwf}) and with $q_0$ in Eq. (\ref{ielslcc}), we transform the former equation at $z_s=z_0+f+z$ into 
\begin{align}
&\psi^{\rm lens}_{\rm IELS}(\Rb,z_0+f+z,t)\approx -\ii \psi_0 f q_0\label{finalstate}\\
&~\times  \ee^{-\ii E^e_0  t/\hbar + \ii q_0(z_0+f+z)}\sum_{\ell=-\infty}^\infty  c_\ell \,\ee^{\ii \ell \omega_{\rm L} (z_0+f+z)/v-\ii \ell \omega_{\rm L}  t},\nonumber 
\end{align}
where now 
\begin{align}
c_\ell=\ee^{-2\pi \ii \ell^2 (z_0+f)/z_{\rm T}}\int_0^\NA d\theta \theta J_\ell[2|\beta(\theta)|]\,\ee^{\ii \ell {\rm arg}\{-\beta(\theta)\}}.\nonumber
\end{align}
In the configuration sketched in Fig. (\ref{Fig4}a), $\beta(\theta)$ is assumed to take constant value $\beta_i$ in the $i$-th of the $M$ concentric sectors of equal normalized area $a$. This directly leads to the simple form $c_\ell=(a/\pi) \ee^{-2\pi \ii \ell^2 d/z_{\rm T}} \sum_{i=1}^M J_\ell (2|\beta_i|)\,\ee^{\ii \ell {\rm arg}\{-\beta_i\}}$ with $d=z_0+f$ used to maximize the fidelity of the light state generated by Eq. (\ref{alphapend1}) and a target state. Because of  the normalization condition, the prefactor in $c_\ell$ does not play any role in the optimization process and thus its output is independent of $a$. Finally, in order to match the form of the electron state in Eq. (\ref{finalstate}) with the one used to arrive at Eq. (\ref{wie}), we absorb the phase $\omega_{\rm L}(z_0+f)/v$ into ${\rm arg}\{-\beta_i\}$.

\subsection{Optimization method}
To find the optimal electron states capable of synthesizing the quantum light states analyzed in this work, the IELS coefficients $\beta_i$ and the propagation distance $d$ are found by employing a random search algorithm combined with a steepest descent method. A maximum number of iterations of $2000$ for the steepest descent together with $3000$ random initial conditions ensured convergence of the results. In Fig. (\ref{FigS2}), we report the optimal coefficients of specific instances shown in Fig. (\ref{Fig4}b-d).

\newpage
\bibliographystyle{apsrev}

\end{document}